\documentclass[preprint,english,aps,prl,groupedaddress,superscriptaddress,floatfix]{revtex4-2}
\usepackage{graphicx,epsfig,units}
\usepackage{xcolor} %colored text, \
\usepackage{soul} %stikeout lines, \st{text}
\usepackage{amsmath,amsfonts,mathrsfs,amsbsy,bm,babel,amssymb}
\usepackage{gensymb}
\usepackage{multibib}
\usepackage{bbm}
\usepackage{xr-hyper}
\makeatletter
\newcommand*{\addFileDependency}[1]{% argument=file name and extension
  \typeout{(#1)}
  \@addtofilelist{#1}
  \IfFileExists{#1}{}{\typeout{No file #1.}}
}
\makeatother

\newcommand*{\myexternaldocument}[1]{%
    \externaldocument{#1}%
    \addFileDependency{#1.tex}%
    \addFileDependency{#1.aux}%
}
\usepackage{xpatch}
\myexternaldocument{EuIn2As2_submitted_ver_SI_Dec_20}

\usepackage[colorlinks=true]{hyperref}  \hypersetup{
	bookmarks=true,         % show bookmarks bar?
	unicode=false,          % non-Latin characters 
	pdftoolbar=true,        % show Acrobat
	pdfmenubar=true,        % show Acrobat 
	pdffitwindow=false,     % window fit to page when opened
	pdfstartview={FitH},    % fits the width of the page to the window
	pdftitle={},    % title
	pdfauthor={},     % author
	pdfsubject={},   % subject of the document
	pdfcreator={},   % creator of the document
	pdfproducer={}, % producer of the document
	pdfkeywords={} {} {}, % list of keywords
	pdfnewwindow=true,      % links in new window
	colorlinks=true,       % false: boxed links; true: colored links
	linkcolor=magenta, %red,          % color of internal links (change box color with linkbordercolor)
	citecolor=blue,        % color of links to bibliography
	filecolor=magenta,      % color of file links
	urlcolor=blue           % color of external links
} 

\newcommand*\diff{\mathop{}\!\mathrm{d}}
\newcommand*\Diff[1]{\mathop{}\!\mathrm{d^#1}}

\newcommand{\cmmnt}[1]{\ignorespaces}
\definecolor{forestgreen}{rgb}{0.13, 0.55, 0.13}

\begin{document}

	\title{Magnetic crystalline-symmetry-protected axion electrodynamics and field-tunable unpinned Dirac cones in EuIn$_\textbf{2}$As$_\textbf{2}$}
	
	\author{S.~X.~M.~Riberolles}
	\email{simon.riberolles@gmail.com}
	\affiliation{Ames Laboratory, Ames, IA, 50011, USA}
	
	\author{T.~V.~Trevisan}
	\affiliation{Ames Laboratory, Ames, IA, 50011, USA}
	\affiliation{Department of Physics and Astronomy, Iowa State University, Ames, IA, 50011, USA}
	
	\author{B.~Kuthanazhi}
	\affiliation{Ames Laboratory, Ames, IA, 50011, USA}
	\affiliation{Department of Physics and Astronomy, Iowa State University, Ames, IA, 50011, USA}
	
	\author{T.~W.~Heitmann}
	\affiliation{University of Missouri Research Reactor, Columbia, MO, 65211, USA}
	
	\author{F.~Ye}
	\affiliation{Oak Ridge National Laboratory, Oak Ridge, TN, 37830, USA}
	
	\author{D.~C.~Johnston}
	\affiliation{Ames Laboratory, Ames, IA, 50011, USA}
	\affiliation{Department of Physics and Astronomy, Iowa State University, Ames, IA, 50011, USA}
	
	\author{S.~L.~Bud'ko}
	\affiliation{Ames Laboratory, Ames, IA, 50011, USA}
	\affiliation{Department of Physics and Astronomy, Iowa State University, Ames, IA, 50011, USA}
	
	\author{D.~H.~Ryan}
	\affiliation{Physics Department and Centre for the Physics of Materials, McGill University, Montreal, Quebec H3A 2T8, Canada}
	
	\author{P.~C.~Canfield}
	\affiliation{Ames Laboratory, Ames, IA, 50011, USA}
	\affiliation{Department of Physics and Astronomy, Iowa State University, Ames, IA, 50011, USA}
	
	\author{A.~Kreyssig}
	\affiliation{Ames Laboratory, Ames, IA, 50011, USA}
	\affiliation{Department of Physics and Astronomy, Iowa State University, Ames, IA, 50011, USA}
	
	\author{A.~Vishwanath}
	\affiliation{Department of Physics and Astronomy, Harvard University, Cambridge, MA, 02138, USA}
	
	\author{R.~J.~McQueeney}
	\affiliation{Ames Laboratory, Ames, IA, 50011, USA}
	\affiliation{Department of Physics and Astronomy, Iowa State University, Ames, IA, 50011, USA}
	
	\author{L.~-L.~Wang}
	\affiliation{Ames Laboratory, Ames, IA, 50011, USA}
	\affiliation{Department of Physics and Astronomy, Iowa State University, Ames, IA, 50011, USA}
	
	\author{P.~P.~Orth}
	\affiliation{Ames Laboratory, Ames, IA, 50011, USA}
	\affiliation{Department of Physics and Astronomy, Iowa State University, Ames, IA, 50011, USA}
	
	\author{B.~G.~Ueland}
	\email{bgueland@ameslab.gov}
	\affiliation{Ames Laboratory, Ames, IA, 50011, USA}

\date{\today}
	\begin{abstract}
	Knowledge of magnetic symmetry is vital for exploiting nontrivial surface states of magnetic topological materials. EuIn$_{2}$As$_{2}$ is an excellent example, as it is predicted to have collinear antiferromagnetic order where the magnetic moment direction determines either a topological-crystalline-insulator phase supporting axion electrodynamics or a higher-order-topological-insulator phase with chiral hinge states.  Here, we use neutron diffraction, symmetry analysis, and density functional theory results to demonstrate that EuIn$_{2}$As$_{2}$ actually exhibits low-symmetry helical antiferromagnetic order which makes it a stoichiometric magnetic topological-crystalline axion insulator protected by the combination of a $180\degree$ rotation and time-reversal symmetries: $C_{2}\times\mathcal{T}=2^{\prime}$. Surfaces protected by $2^{\prime}$ are expected to have an exotic gapless Dirac cone which is unpinned to specific crystal momenta. All other surfaces have gapped Dirac cones and exhibit half-integer quantum anomalous Hall conductivity. We predict that the direction of a modest applied magnetic field of $\mu_{0}H\approx1$ to $2$~T can tune between gapless and gapped surface states. 
	
	\end{abstract}
	\maketitle	
	
	%%%%%%%%%%%%%%%%%%%%%%%%%%%%%%%%%%%%%%%%%%%%%%
	%% The main text: 
	%% The text is limited to 3000 words
	%%%%%%%%%%%%%%%%%%%%%%%%%%%%%%%%%%%%%%%%%%%%%%
\section{Introduction}
	Electrons attaining a nontrivial Berry phase due to symmetry-protected features in the electronic band structure \cite{Ref_1,Ref_2, Ref_3,Ref_4} and/or the presence of noncoplanar magnetic order \cite{Ref_5,Ref_6,Ref_7} can lead to astonishing topological physical properties such as dissipationless chiral-charge transport, quantum anomalous Hall (QAH) effect, and axion electrodynamics  \cite{Ref_4,Ref_8}. Whereas topological-crystalline insulators (TCIs) are broadly defined as insulators with nontrivial topological properties protected by crystalline symmetry,  magnetic TCIs offer the possibility of tuning topological properties via manipulating the magnetic order \cite{Ref_4}.  Indeed, much theoretical effort is focused on predicting magnetic crystalline materials with nontrivial topological states by considering symmetries associated with the magnetic space groups (MSGs) describing their magnetic order \cite{Ref_9,Ref_10,Ref_11}.  A database with predictions for nontrivial topological band structures based on MSG symmetries provides important guidance towards finding new magnetic topological materials using high-throughput \emph{ab initio} studies \cite{Ref_11}.  However, an often limiting bottleneck is detailed knowledge of a candidate's intrinsic magnetic order, which is difficult to predict theoretically. Determining such order can be a subtle task requiring significant experimental effort.
    
    QAH and axion insulators (AXIs) are particularly attractive topological states as the former manifests quantized Hall conductivity in the absence of an applied magnetic field, and the latter shares similarities with the axion particle in quantum chromodynamics \cite{Ref_12}. AXIs exhibit the topological magnetoelectric (TME) effect for which an applied electric field $\mathbf{E}$ induces a parallel magnetization $\mathbf{M}$ or a magnetic field $\mathbf{H}$ induces a parallel electric polarization \cite{Ref_4}. An AXI requires that the axion angle $\theta$ in the action of axion electrodynamics $S_{\theta} = \theta\frac{e^2}{4 \pi^2} \int \diff t \Diff3 r \mathbf{E}\cdot \mathbf{B}$ is $\theta=\pi$, which leads to the presence of half-integer QAH-type conductivity on insulating surfaces \cite{Ref_13,Ref_10}.  Here, $\mathbf{B}=\mu_{0}(\mathbf{H}+\mathbf{M})$ is the magnetic induction, $\mu_{0}$ is the permeability of free space, and $e$ is the electron charge. $\theta$ is quantized to zero or $\pi$ in the presence of either time-reversal $\mathcal{T}$ or inversion $\mathcal{I}$ symmetry, but also any other symmetry operation that reverses an odd number of space-time coordinates~ \cite{Ref_14}.

\begin{figure} %%% CORELLI 2D%%%
		\centering
		\vspace{0mm}
		\includegraphics[width=1.0\textwidth,trim=0.0cm 0.0cm 0.0cm 0.0cm]{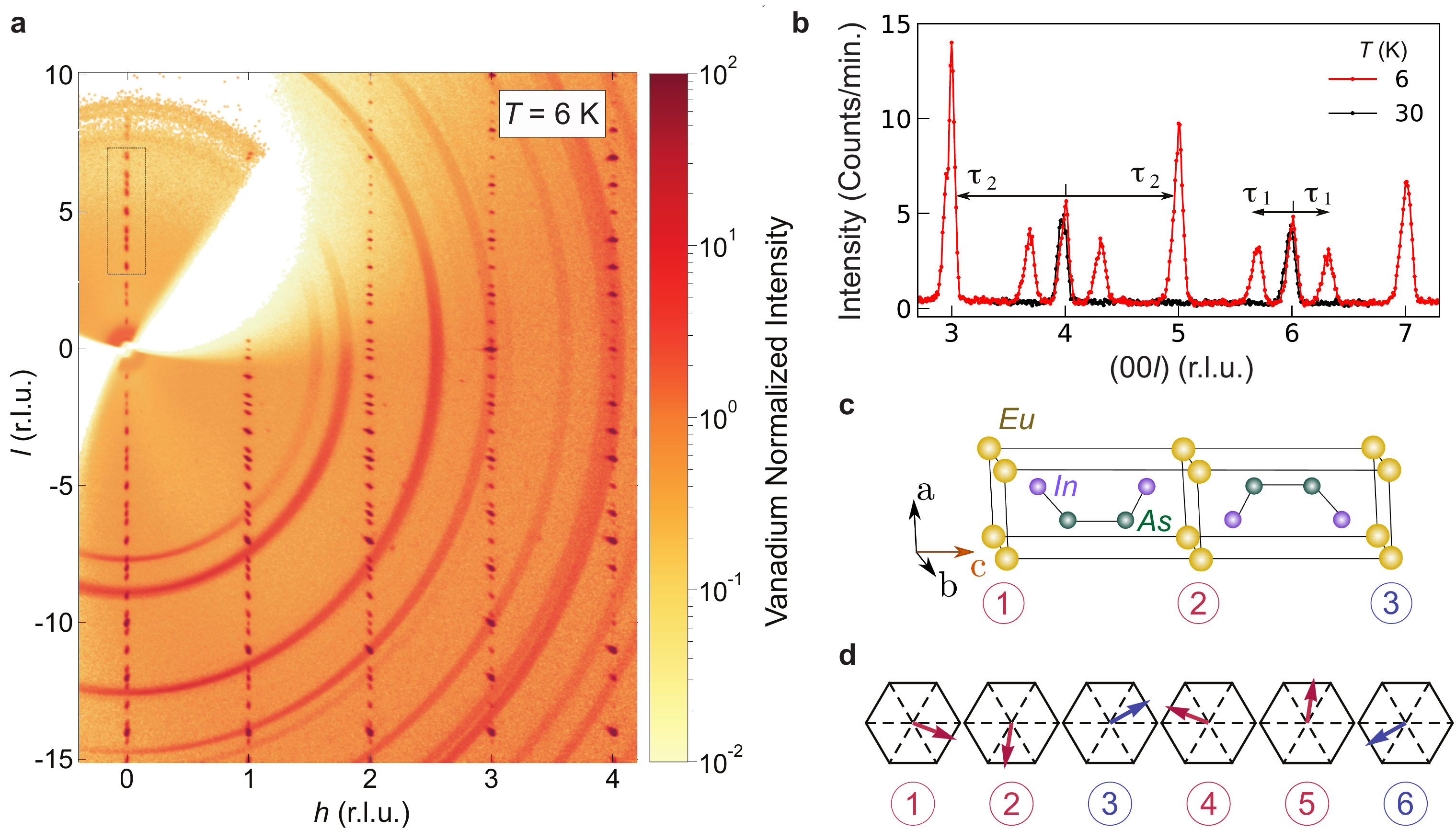}
		\vspace{0mm}
		\caption {\textbf{Neutron diffraction determination of the broken-helix magnetic order.} 
			\textbf{a}, Data for the $(h0l)$ reciprocal-lattice plane measured at a temperature of $T=6$~K with the CORELLI spectrometer \cite{Ref_19}.  Bragg peaks from the single-crystal sample appear as dots, whereas Bragg peaks due to the polycrystalline Al sample holder appear as rings centered at $0$. ~\textbf{b}, Scans along the $(00l)$ reciprocal-lattice direction for $6$~K (red) and $30$~K (black) made with the TRIAX instrument.  The $30$~K data show Bragg peaks solely due to the chemical lattice, and the two sets of magnetic Bragg peaks appearing in the $6$~K data are labeled by the antiferromagnetic propagation vectors $\bm{\tau}_{1} = (0,0,\tau_{1z})$, with $\tau_{1z}=0.303(1)$, and $\bm{\tau}_{2} = (0,0,1)$ \cite{Ref_20}.  \textbf{c}, Chemical unit cell of EuIn$_{2}$As$_{2}$ where the numbers correspond to labeling in panel d. \textbf{d}, Layer-by-layer diagram of the $6$~K broken-helix order using the notation of the hexagonal chemical unit cell. The magnetic unit cell is actually orthorhombic and is tripled along $\mathbf{c}$ with the $\mathbf{a}_{\text{ortho}}$ and $\mathbf{b}_{\text{ortho}}$ orthorhombic unit-cell axes lying along  $[\bar{1}10]$ and $-[110]$, respectively. Each Eu layer is ferromagnetically aligned with the ordered magnetic moments lying in the $\mathbf{ab}$ plane. The two symmetry-inequivalent magnetic Eu sites in the magnetic unit cell are colored red (layers $1$, $2$, $4$ and $5$) and blue (layers $3$ and $6$).  Moments in the blue layers are constrained by symmetry to lie along $[\bar{1}10]$.}
		\label{Diff}
	\end{figure}

	Hexagonal EuIn$_{2}$As$_{2}$ [space goupe $P6_{3}/mmc$ (No.~$194$) with lattice parameters $a=4.178(3)$~\AA\ and $c=17.75(2)$~\AA] is a magnetic TCI built of alternating Eu and In$_2$As$_2$ layers stacked along $\mathbf{c}$ as shown in Fig.~\ref{Diff}c \cite{Ref_15,Ref_16}.  The magnetic  Eu$^{2+}$ (spin $S= \frac{7}{2}$) layers undergo antiferromagnetic (AF) ordering at a N\'{e}el temperature of $T_{\text{N}}\approx18$~K \cite{Ref_17} which density functional theory (DFT) calculations predict to be A-type \cite{Ref_18}.  A-type order is collinear, with the ordered Eu magnetic moments $\bm{\mu}$ ferromagnetically aligning in each layer and the layers stacking AF along $\mathbf{c}$.  Theory predicts that depending on the orientation of $\bm{\mu}$, the A-type order leads to an AXI that is either a TCI with some gapless surfaces or a higher-order topological insulator with chiral-hinge states \cite{Ref_18,Ref_17}. Attractively, the ordered Eu moments may influence topological fermions in In$_{2}$As$_{2}$ layers, providing a path for \emph{in situ} control of band topology.

    Below we detail our discovery of low-symmetry broken-helix magnetic order in EuIn$_{2}$As$_{2}$ using single-crystal neutron diffraction.  We find that the broken-helix order has $2^{\prime}$ symmetry elements along specific crystalline axes that lead to the emergence of an AXI in the absence of $\mathcal{I}$ and $\mathcal{T}$, both of which are broken by the magnetic ordering. The $2^{\prime}$ symmetry element denotes the product of a two-fold rotation ($C_{2}$) and the $\mathcal{T}$ operation: $2^{\prime}=C_2\times\mathcal{T}$.  $C_2$ rotates the lattice and the spins around the axis by $\pi$ and  $\mathcal{T}$ then reverses the direction of the spins. As shown in detail in Supplementary Figs.~\ref{fig:symmetry} and \ref{fig:symmetry_broken}, the $2^{\prime}$ transformations leave the broken-helix order invariant and protect gapless surface Dirac cones that are not pinned to time-reversal-invariant momenta (TRIM). Surfaces not associated with a $2^{\prime}$ axis have gapped Dirac cones and exhibit half-integer QAH-type conductivity. Our symmetry analyses and magnetization data predict that the surface states are highly tunable by a modest field of  $\mu_{0}H\approx1$ to $2$~T which is strong enough to polarize the magnetic moments along its direction.  This induces a gap and, depending on the field's direction,  creates new unpinned gapless states on previously gapped surfaces.
    
\section{Results}
\textbf{Neutron diffraction determination of the antiferromagnetic order.} Figure~\ref{Diff}a shows neutron diffraction data for  the $(h0l)$ reciprocal-lattice plane taken below $T_{\text{N}}\approx18$~K at $T=6$~K, and Fig.~\ref{Diff}b shows data along the $(00l)$ direction taken above and below $T_{\text{N}}$ at $30$ and $6$~K.  Nuclear Bragg peaks occur in Fig.~\ref{Diff}b at $l=2n$ positions for both temperatures whereas magnetic Bragg peaks only appear in the $6$~K data.   Two AF propagation vectors define the locations of the magnetic Bragg peaks:  ($1$) peaks at $l=2n \pm \tau_{1z}$ positions correspond to $\bm{\tau}_{1}=(0,0,\tau_{1z})$ with $\tau_{1z}=0.303(1)$; ($2$) nuclear-forbidden peaks at $l=2n+1$ positions in Fig.~\ref{Diff}b correspond to $\bm{\tau}_{2}=(0,0,1)$ \cite{Ref_20}. Within the $(h0l)$ reciprocal-lattice plane, magnetic Bragg peaks matching $\bm{\tau}_{2}$ overlap with nuclear Bragg peaks with odd values of $l$ when the conditions $h=3n+1$ or $h=3n+2$ ($n=\text{integer}$) are satisfied.   The predicted A-type order would be consistent with magnetic Bragg peaks appearing only at positions corresponding to $\bm{\tau}_{2}$.  However, the existence of additional magnetic Bragg peaks at $\bm{\tau}_{1}$ reveals the presence of more complex helical or spin-density-wave type (itinerant) AF order. Our measurements cannot distinguish between these two types of order, but our DFT calculations reveal that the Eu $4f$ bands are located well below the Fermi level which is more supportive of local-moment helical order. 

Using magnetic symmetry analysis and single-crystal refinements to our data from the TRIAX triple-axis spectrometer and the CORELLI time-of-flight spectrometer, we discover that EuIn$_{2}$As$_{2}$ has the complex broken-helix AF order illustrated in Fig.~\ref{Diff}d and Supplementary Figure~\ref{bH_order}.  It consists of ferromagnetically-aligned layers with $\bm{\mu}\perp\mathbf{c}$ that are helically stacked along $\mathbf{c}$.  We find that the orthorhombic MSG $C2^{\prime}2^{\prime}2_{1}$ (No.~$20.33$) describes the symmetry, but for continuity we refer to directions with respect to the hexagonal chemical unit cell.  We find $\mu=6.0(3)~\mu_{\text{B}}/\text{Eu}$ at $T=6$~K, which is smaller than the expected value of $7.0~\mu_{\text{B}}/\text{Eu}$ and the saturation moment of $\mu_{\text{sat}}=7.00(6)~\mu_{\text{B}}$ shown below and in  Supplementary Figure~\ref{characterization_2}.  We note that neutron diffraction studies of other Eu containing compounds report less than $7.0~\mu_{\text{B}}/\text{Eu}$~\cite{Ref_21,Ref_22}, and, as shown below, the ordered moment is still increasing with decreasing temperature below $6$~K.   Details of the corrections performed to account for the strong thermal neutron absorption of Eu are given in the Methods Section and Supplementary Note~1.
	
To analyze the broken-helix order, we assume a tripling of the chemical unit cell since $\tau_{z}=0.303(1)\approx\frac{1}{3}$ and the relatively small incommensurability results in maximum disagreement only at the $33$rd Eu layer.  Referring to Fig.~\ref{Diff}d, we designate the two Eu crystallographic sites in $C2^{\prime}2^{\prime}2_{1}$ as red and blue.  This in turn indicates red and blue ferromagnetically-aligned layers, and allows for defining the helix turn angles $\phi_{\text{rr}}$ and $\phi_{\text{rb}}$ between successive red-red and red-blue layers, respectively. The MSG symmetry dictates that $\phi_{\text{rr}} + 2\phi_{\text{rb}} =180\degree$ and constrains magnetic moments in the blue layers to lie along $[\bar{1}10]$. Our refinements find that $\phi_{\text{rb}} = 127(3)\degree$ at $T=6$~K, and more details of the refinements are given in Supplementary Figures~\ref{MSG_comparison}--\ref{TRIAX_order_para} and Supplementary Tables~\ref{Table_MSG_broken_helix}--\ref{TRIAX_data_table}.

	\begin{figure} %%% spin_angle_phase_diag%%%
		\centering
		\includegraphics[width=0.9\textwidth,trim=0.0cm 0.0cm 0.0cm 0.0cm]{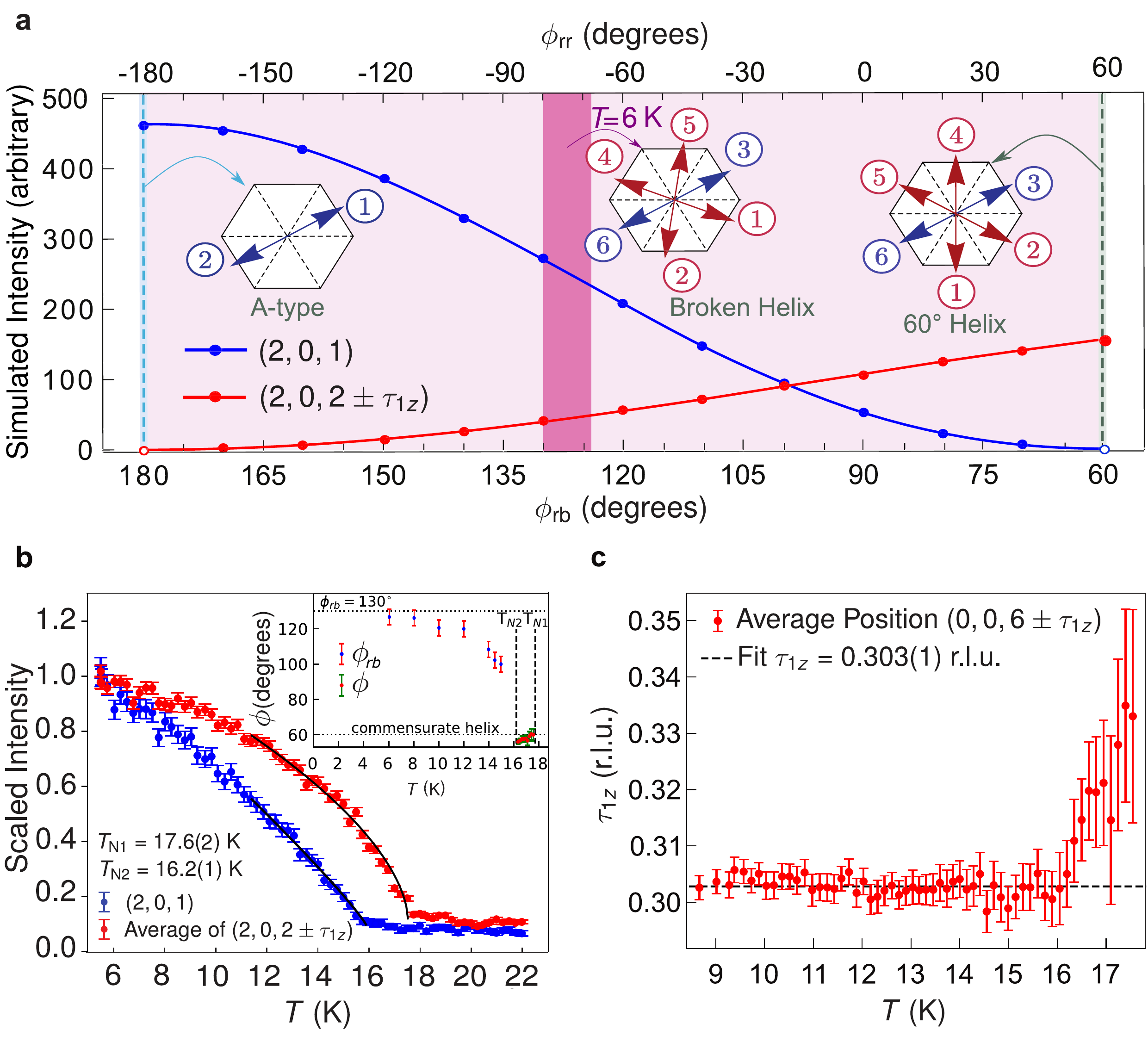}
		\caption{\textbf{Details of the broken-helix magnetic order.} \textbf{a}, Simulations of the magnetic neutron diffraction intensity for the $(2,0,1)$ Bragg peak and the average of the $(2,0,2) \pm \bm{\tau}_{1}$ Bragg peaks as functions of the broken-helix turn angles $\phi_{\text{rb}}$ and $\phi_{\text{rr}}$.  \textbf{b}, Temperature dependencies of the integrated intensity of the $(2,0,1)$ Bragg peak and the average integrated intensity for the $(2,0,2) \pm \bm{\tau}_{1}$ magnetic Bragg peaks measured on the CORELLI spectrometer and scaled to $1$ at $T=6$~K. The inset displays the helix turn angles $\phi$ in the pure $60\degree$-helix phase ($T_{\text{N}2}<T\leq T_{\text{N}1}$) and $\phi_{\text{rb}}$ for the broken-helix phase ($T\leq T_{\text{N}2}$).  \textbf{c}, Temperature dependence of $\bm{\tau}_{1}=(0,0,\tau_{1z})$ from the average of the positions of the $(0,0,6)\pm\bm{\tau}_{1}$ magnetic Bragg peaks. In b, $\phi(T)$ is calculated from the temperature evolution of $\tau_{1z}$.  $\phi_{\text{rb}}(T)$ is calculated from the temperature dependence of the ratio of the integrated intensity of the $(0,0,5)$ magnetic Bragg peak to that of the $(006)-\tau_{1}$ magnetic Bragg peak.  These data are from measurements made on the TRIAX spectrometer and are shown in  Supplementary Figure~\ref{TRIAX_order_para}.}
		\label{fig_bloc}
	\end{figure}
	
We simulate in Fig.~\ref{fig_bloc}a the integrated intensities for representative magnetic Bragg peaks for different values of $\phi_{\text{rb}}$ using the determined MSG and $\bm{\mu}\perp\mathbf{c}$.  Magnetic Bragg peaks corresponding to $\bm{\tau}_2$  disappear as $\phi_{\text{rb}}\rightarrow60\degree$ (pure $60\degree$-helix order), whereas magnetic Bragg peaks matching $\bm{\tau}_1$ disappear as $\phi_{\text{rb}}\rightarrow180\degree$ (A-type order).  We find that the pure $60\degree$-helix order can be described by the higher-symmetry MSG $P6_12^{\prime}2^{\prime}$ (No.~$178.159$) and that the A-type order can be described by MSG $Cm^{\prime}c^{\prime}m$ (No.~$63.462$). The darker shaded area in Fig.~\ref{fig_bloc}a indicates the experimental value of $\phi_{\text{rb}}$ at $T=6$~K.
	
Figures~\ref{fig_bloc}b and \ref{fig_bloc}c reveal the temperature evolution of the AF order.  Figure~\ref{fig_bloc}b displays the measured integrated intensities of the $(2,0,2)\pm \bm{\tau}_1$ and $(2,0,1)$ magnetic Bragg peaks scaled to $1$ at $T=6$~K.  The curves suggest that $\mu$ continues to increase with decreasing temperature below $6$~K. Two transitions are evident: the incommensurate magnetic Bragg peaks corresponding to $\bm{\tau}_{1}$ with $\tau_{1z}\approx\frac{1}{3}$ emerge at $T_{\text{N}1}=17.6(2)$~K and commensurate magnetic Bragg peaks corresponding to $\bm{\tau}_{2}$ appear at $T_{\text{N}2}=16.2(1)$~K.    These transitions also appear in the magnetic susceptibility, resistance, and $^{151}$Eu M\"{o}ssbauer data shown in  Supplementary Figure~\ref{characterization_1}.  Figure~\ref{fig_bloc}c displays the temperature dependence of $\bm{\tau}_1$ which indicates that pure $60\degree$-helix order first emerges at $T_{\text{N}1}$, and $\tau_{1z}$ decreases upon cooling until $T_{\text{N}2}$ where broken-helix order emerges.  This sequence follows the rightmost path in the group-subgroup chart in  Supplementary Figure~\ref{MSG_tree} which traces second-order magnetic transitions from the paramagnetic state to pure $60\degree$-helix order to broken-helix order.

The inset to Fig.~\ref{fig_bloc}b shows the evolution of the helix turn angle $\phi$ below $T_{\text{N}1}$. $\phi\approx60\degree$ at $T_{\text{N}1}$ and slightly decreases upon cooling for $T_{\text{N}2}<T<T_{\text{N}1}$.  The small change from $60\degree$ reflects the change in $\tau_{1z}$. For $T<T_{\text{N}2}$, the broken-helix order emerges and the inset to Fig.~\ref{fig_bloc}b plots $\phi=\phi_{\text{rb}}$, which is calculated from data shown in Supplementary Figure~\ref{TRIAX_order_para}. $\phi_{\text{rb}}$ rapidly increases immediately below $T_{\text{N}2}$ and approaches $\phi_{\text{rb}}=130\degree$ as $T\rightarrow0$~K. The temperature evolution of $\phi_{\text{rb}}$ below $T_{\text{N}2}$, while maintaining $\tau_{1z}=0.303(1)$, agrees with the calculations in Fig.~\ref{fig_bloc}a showing that broken-helix order persists over a range of $\phi_{\text{rb}}$.
    
	\begin{figure} %%Band Structure
		\centering
		\vspace{0mm}
		\includegraphics[width=1.0\textwidth]{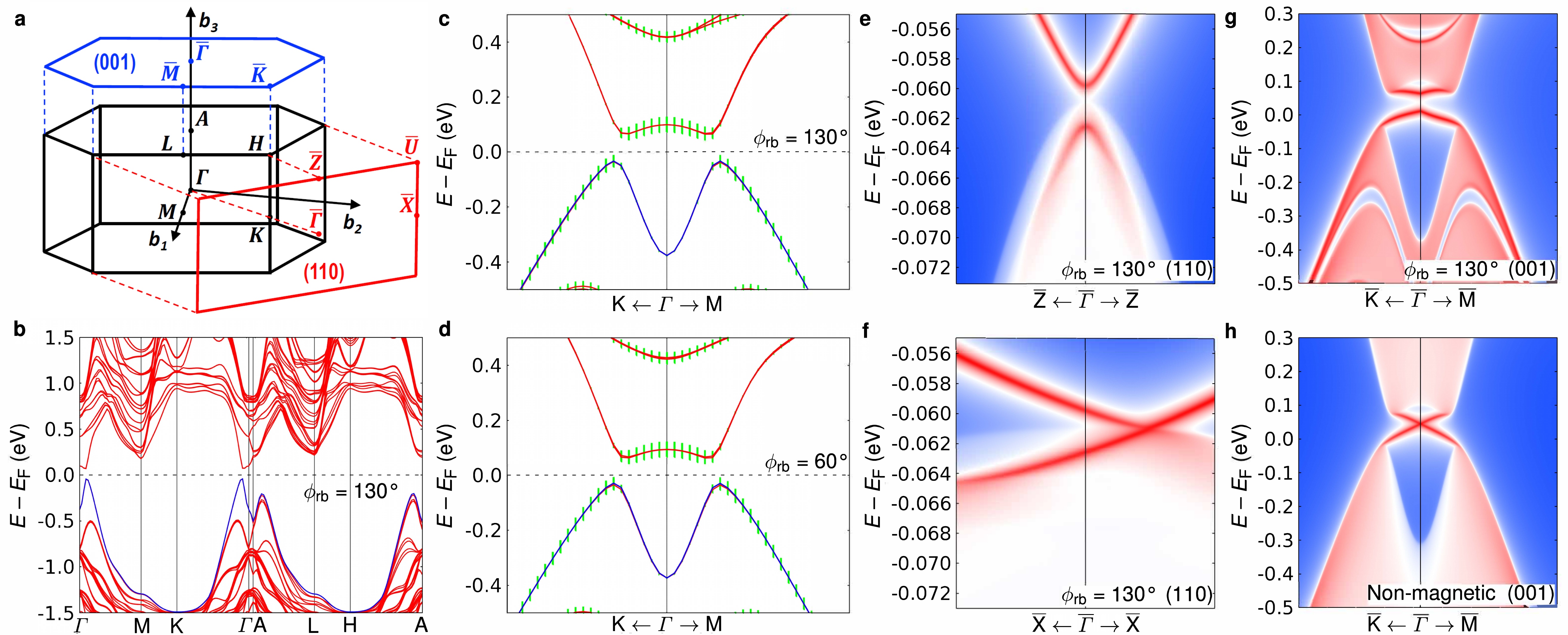}
		\caption{\textbf{Bulk and surface electronic band structures for various magnetic ground states.}   \textbf{a}, Hexagonal Brillouin zone and the projected $(110)$ and $(001)$ surface-Brillouin zones with high-symmetry points (capital letters) and reciprocal-lattice axes ($\mathbf{b}_{1}=\mathbf{a^*}$, $\mathbf{b}_{2}=\mathbf{b^*}$, and $\mathbf{b}_{3}=\mathbf{c^*}$ ) indicated.  \textbf{b}, Bulk band structure along high-symmetry paths for  broken-helix order with $\phi_{\text{rb}} =130\degree$.  $E_{\text{F}}$ is the Fermi energy.  \textbf{c,d}, Views of the minimal gap region along the $\mathrm{K}$-$\Gamma$-$\mathrm{M}$ direction for  $\phi_{\text{rb}} =130\degree$ (broken-helix order) (c)  and $\phi_{\text{rb}}=60\degree$ (pure $60\degree$-helix order) (d) . The top valence band according to simple band filling is indicated in blue and the rest of the bands are colored red. Green vertical bars show As $4p_z$ orbital character and indicate band inversion near the minimal gap.  \textbf{e--h} Surface bands for the $(110)$ (e,f) and $(001)$ (g,h) projected surfaces along high-symmetry lines for $\phi_{\text{rb}} =130\degree$ (e,f,g).  A Dirac cone exists on each surface, but the presence of a gap and whether or not the cone is pinned to a time-reversal-invariant momentum (TRIM) depends on symmetry:  a gapless Dirac cone unpinned from a TRIM occurs for the $(110)$ surface preserving the $2^{\prime}$ symmetry axis.  On the other hand, the $(001)$ surface (g) has a gapped Dirac cone at $\bar{\Gamma}$ due to the absence of a $2^{\prime}$ axis and any other symmetry that reverses an odd number of space-time coordinates. The surfaces with gapped Dirac cones support half-integer quantum-anomalous-Hall-type conductivity. The nonmagnetic calculation for the $(001)$ surface (h) has a gapless Dirac cone pinned to $\bar{\Gamma}$. Note that panels e and f are zoomed in around the bottom of the band gap to show the details of the surface Dirac cone. $E_{\text{F}}$ is inside the gap and the surface Dirac point is at $E_{\text{F}}-0.061$~eV, very close to the top of valence band projection.   More details of the surface band structure are found in Supplementary Fig.~\ref{DFT_SI}g.}
		\label{band_structure}
	\end{figure}

  \textbf{Density functional theory calculations and symmetry analyses.} Having established and described the emergence of complex helical order, we now discuss and compare the impacts of broken-helix, pure $60\degree$-helix, and A-type order on the electronic band structure and topology.  First, similar to previous reports \cite{Ref_18}, our DFT calculations and symmetry analyses find that A-type order creates an AXI state with inverted bulk bands near $\Gamma$ and an electronic band gap of $\approx 100$~meV. Looking at the phase diagram in Fig.~\ref{fig_bloc}a, we realize that by adiabatic continuation the magnetic symmetry can be lowered away from the pure $60\degree$-helix or A-type ordered states by varying $\phi_{\text{rb}}$ while maintaining the constraints that $\phi_{\text{rr}} + 2\phi_{\text{rb}} =180\degree$ and ordered moments in the blue layers lie along $[\bar{1}10]$.  For example, starting from A-type order, reducing $\phi_{\text{rb}}$ from $180\degree$ immediately breaks the mirror and $\mathcal{I}$ symmetries present in MSG $Cm^{\prime}c^{\prime}m$.
  
Using this adiabatic continuation approach, we find via DFT calculations that a full band gap persists for any value of $\phi_{\text{rb}}$. Figure~\ref{band_structure}a shows a diagram of the Brillouin zone and Fig.~\ref{band_structure}b shows the calculated electronic bands for $\phi_{\text{rb}}=130\degree$.  Figure~\ref{band_structure}c focuses on the gap region near $\Gamma$ where band inversion is highlighted by green bars denoting As $4p_z$ orbital character, and  Fig.~\ref{band_structure}d demonstrates that the inverted gap remains for pure $60\degree$-helix order. Supplementary Figure~\ref{DFT_SI} shows that this result holds for other values of $\phi_{\text{rb}}$ spanning $180\degree\leq\phi_{\text{rb}}\leq60\degree$. Thus, the inverted gap remains across all three magnetic orders connected by the internal parameter $\phi_{\text{rb}}$ and indicates that the topological phase is robust to changes to $\phi_{\text{rb}}$. In contrast, we find that bands cross the Fermi level for the case of ferromagnetic order as shown in Supplementary Figure~\ref{DFT_SI}.  This opens the possibility of controlling the band structure by applying a magnetic field strong enough to align $\bm{\mu}$ along a single direction.  This point is discussed further below.

\section{Discussion}   
  With these results in hand, we now address why EuIn$_{2}$As$_{2}$ is an AXI in its magnetically-ordered phases despite the absence of $\mathcal{I}$.  We determine $\theta=\pi$ in $S_{\theta}$ for the broken-helix order by finding that $\theta = \pi$ in the presence of $\mathcal{I}$ for $\phi_{\text{rb}}=180\degree$ (A-type order) via calculating the parity-based symmetry indicator $\mathbb{Z}_4 = \frac12 \sum_{k \in \text{TRIMs}} (n^+_k - n^-_k) \, \text{mod} \, 4=2$ \cite{Ref_10,Ref_23,Ref_11,Ref_18}. Here, $n^\pm_k \geq 0$ denotes the total number of filled bands with parity eigenvalue $\pm 1$ at TRIM $k$.  Since the band gap remains for $60\degree\leq\phi_{\text{rb}} \leq180\degree$,  adiabatic continuity ensures that $\theta=\pi$ for all three magnetic structures, and, remarkably, the entire family of A-type, broken-, and pure $60\degree$-helix states are AXIs as long as the Fermi energy resides in the band gap. It is the $2^{\prime}$ symmetry operator along $[110]$ and $[\bar{1}10]$  for the broken-helix order and along $[110]$ and 
  $[100]$ for the pure $60\degree$-helix order that permits an AXI in the absence of $\mathcal{I}$ \cite{Ref_14}.
  
    The presence of $2^\prime$ symmetry axes for the broken-helix order enforces an odd number of gapless unpinned surface Dirac cones on the $(\bar{1}10)$ and $(110)$ surfaces.  More specifically, the unpinned gapless Dirac cones occur on surfaces perpendicular to the $2^\prime$ axes. These exotic topological surface states appear in the presence of $2^\prime$ because it reverses an odd number of space-time coordinates \cite{Ref_14}. Figures~\ref{band_structure}e and \ref{band_structure}f show the surface band structure for $(110)$ wherein a gapless Dirac cone is seen which is offset from $\bar{\Gamma}$ along $\bar{\Gamma}$-$\bar{\mathrm{X}}$.  This result contrasts with the gapped Dirac cone at $\bar{\Gamma}$ on the $(001)$ surface shown in Fig.~\ref{band_structure}g.  For the case of no magnetic order,  Fig.~\ref{band_structure}h shows a gapless Dirac cone on the $(001)$ surface pinned to $\bar{\Gamma}$.  These calculations did not include the Eu $4f$ orbitals. We find that $\mathbb{Z}_{2}=1$ which signals a strong topological insulator, rather than an AXI phase, for the non-magnetic state.
    
	\begin{figure} %%Field Polarized
		\centering
		\vspace{0mm}
		\includegraphics[width=1.0\textwidth]{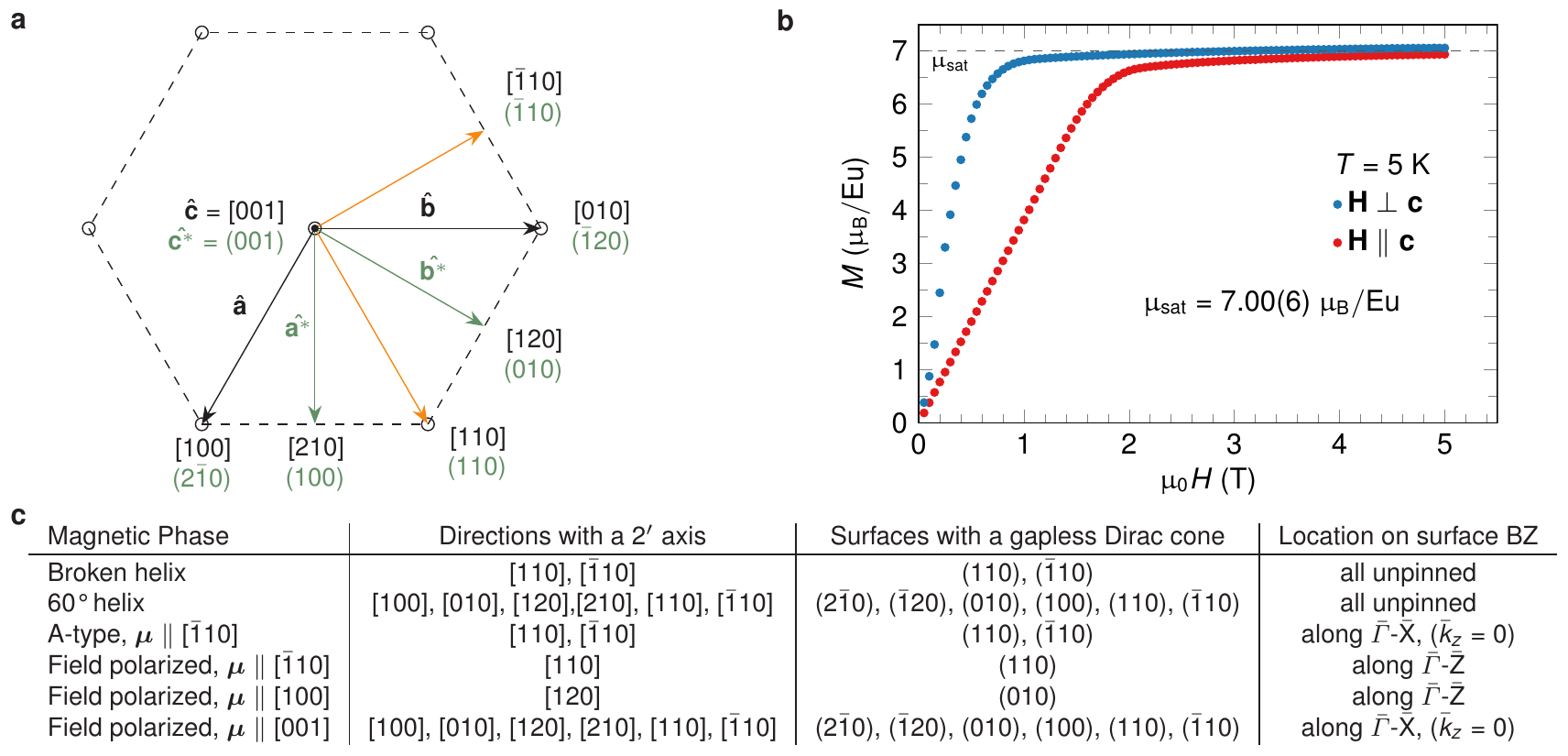}
		\caption{\textbf{Tuning surface Dirac cones via an applied magnetic field. }   \textbf{a}, Diagram showing real-space (denoted by brackets) and reciprocal-space (denoted by parentheses) directions for the $\mathbf{ab}$ plane.  The magnetic orthorhombic unit cells for broken-helix and A-type orders discussed in the main text have their $\mathbf{a}_{\text{orth}}$ and $\mathbf{b}_{\text{orth}}$ axes along $[\bar{1}10]$ and $-[110]$, respectively. \textbf{b}, Magnetization versus magnetic field curves for a temperature of $T=5$~K and the field applied either perpendicular or parallel to $\mathbf{c}$.   \textbf{c}, A table indicating the $2^\prime$ axis/axes, the surface(s) containing a gapless Dirac cone, and the location of the gapless Dirac cone on the projected surface Brillouin Zone(s) (BZ) for various zero-field and field-polarized magnetic states.  The direction of the ordered magnetic moment $\bm{\mu}$ is indicated where appropriate.  The results for the broken-helix and $60\degree$-helix phases assume that the blue Eu layers lie along $[\bar{1}10]$.}
		\label{Field_Pol}
	\end{figure}
    
     The AXI state protected by $2^{\prime}$ symmetry is different from the AXI phase protected by a combination of $\mathcal{T}$ and a translation recently proposed for MnBi$_2$Te$_4$~\cite{Ref_24}. In MnBi$_2$Te$_4$, this symmetry is only unbroken on surfaces that contain the translation vector, where it leads to gapless Dirac cones that are pinned to TRIM. The Dirac cones are gapped on all other surfaces. In contrast, in EuIn$_2$As$_2$ the AXI state is protected by $2^{\prime}$ symmetry, leading to unpinned Dirac surface states whenever the $2^{\prime}$ axis is normal to the surface and gapped surface states appear on all other surfaces.
    
    The different topological surface states resulting from the presence or absence of $2^{\prime}$ symmetry leads to differing physical properties.  To facilitate this discussion, Fig.~\ref{Field_Pol}a shows the relationships between real-space and reciprocal-lattice directions, and the direction(s) of the $2^{\prime}$ axis/axes are listed in Fig.~\ref{Field_Pol}c for various magnetic states.  For the broken-helix order, $(001)$ and other gapped surfaces without a $2^{\prime}$ axis exhibit half-integer QAH-type conductivity \cite{Ref_4}. On the other hand, similar to a $3$D topological insulator, the gapless surface Dirac cones on $(110)$ and $(\bar{1}10)$ support dissipationless charge transport where sweeping $E_{\text{F}}$ through the Dirac point switches the handedness of the chirality associated with the itinerant charge.  Similar analyses of the surface states can be performed for the pure $60\degree$-helix and A-type order phases.  In particular, Fig.~\ref{Field_Pol}c shows that the pure $60\degree$-helix phase supports multiple surfaces with an unpinned gapless Dirac cone.   The consequences of the gapless Dirac cones being exotically unpinned from TRIM, which require absence of Kramer's degeneracy, however, lacks investigation.
   
   Figure~\ref{Field_Pol}c also illustrates that the surface states are highly tunable by an external magnetic field, and Fig.~\ref{Field_Pol}b shows that modest field strengths of $\mu_{0}H \approx 1$ to $2$~T are sufficient to polarize the Eu magnetic moments along any direction at $T=5$~K.  For example, the row in Fig.~\ref{Field_Pol}c corresponding to broken-helix order indicates that a gapless completely unpinned Dirac cone exists only on the $(\bar{1}10)$ and $(110)$ surfaces, and, therefore, all of the other surfaces have a gapped Dirac cone.  For a field-polarized state with $\bm{\mu}\parallel[100]$, Fig.~\ref{Field_Pol}c indicates that only the $(010)$ surface has a gapless surface Dirac cone and its node is constrained to lie along $\bar{\Gamma}$-$\bar{\mathrm{Z}}$. Thus, a field provides an effective means for switching by inducing a gap in previously gapless surface states while eliminating the gap in previously gapped surface states.  Further, the position of the node of the gapless Dirac cone can change from being completely unpinned to being constrained to a line. Figure~\ref{Field_Pol}c also highlights the direct correlation between the existence of a $2^{\prime}$ axis and gapless surface states.  More details of our symmetry analyses are given in Supplementary Table~\ref{tab:my-table} and in the associated discussion in Supplementary Note~5.
   
   Finally, a true AXI state only occurs for insulating compounds, but resistance and  angle-resolved-electron-photoemission data for EuIn$_2$As$_2$ are consistent with it being a slightly hole-doped compensated semimetal \cite{Ref_17,Ref_25}, see Supplementary Figure~\ref{characterization_1}c. Thus, $E_{\text{F}}$ needs to be shifted into the bulk gap for an AXI to be observed.  Fortunately, the robust quantization of $\theta = \pi$ with respect to changes of $\phi_{\text{rb}}$ suggests that small perturbations to the magnetic order potentially caused via tuning $E_{\text{F}}$ will not destroy the symmetry leading to an AXI state. Further, the fact that $\bm{\tau}_1$ is not exactly $\frac{1}{3}$ does not invalidate our analysis, because the slight incommensurability should enter into the calculation of $\theta$ as only a small perturbation.  Investigations into growing thin films may permit gating or strain to shift the chemical potential, and growth of films on magnetic substrates may allow for simultaneous gating while facilitating magnetic-field control of the topological surface states.
   
    To conclude, we have established that complex helical magnetic order emerges in EuIn$_{2}$As$_{2}$ upon cooling below $T_{\text{N}1}=17.6(2)$~K (pure $60\degree$-helix order) and $T_{\text{N}2}=16.2(1)$~K  (broken-helix order).  Both of these helical orders create $2^{\prime}$ symmetry axes that protect an AXI state in the absence of $\mathcal{I}$.  Our DFT calculations and symmetry analyses show that the band topology is robust to changes of the helical turn angles, such that the magnetic order can change from A-type, to broken-helix, to pure $60\degree$-helix order while maintaining the same topological phase.  We further find that the existence of $2^{\prime}$ symmetry axes protects gapless Dirac cones on specific surfaces which are unpinned from TRIM and that the other surfaces possess gapped Dirac cones and exhibit QAH-type conductivity.  Our magnetization results show that only a moderate magnetic field of $\mu_{0}H\approx1$ to $2$~T is necessary for tuning the direction of the magnetic moment, which our symmetry analyses show can choose the surfaces supporting gapless unpinned Dirac cones. This opens up the possibility for novel device development as well as further research into the physical properties associated with unpinned surface Dirac cones.  For example, experiments exploring if these exotic surface states can be moved in the surface Brillouin zone via pressure or strain may yield insight into basic material interactions and reveal unexpected  properties.  The intrinsic low-symmetry broken-helix magnetic order of EuIn$_{2}$As$_{2}$ offers an excellent opportunity to study and tune  exotic surface states in a magnetic topological-crystalline insulator.

	\section{Methods}
	\textbf{Sample synthesis.} Single crystals of EuIn$_2$As$_2$ were grown using a flux method similar to Ref.~\cite{Ref_15} and found to be single phase via powder x-ray diffraction. An initial composition of Eu:In:As~$=3:36:9$ was weighed out and packed in fritted alumina crucibles \cite{Ref_26}, followed by sealing in a fused silica tube. The prepared ampoule was first heated up to $300\degree$C over $2$ hours and held there for an hour, then heated up to $580\degree$C over $3$ hours and held there for $2$ hours, and finally heated up to $900\degree$C over $10$ hours followed by a $2$ hour dwell. The intermediate dwells were to ensure maximum dissolving and incorporation of the volatile elements into the melt. The final dwell was followed by a $48$ hour cool down to $770\degree$C, at which point excess flux was decanted using a centrifuge. This process yielded plate-like crystals of EuIn$_2$As$_2$ with masses of a few milligrams.
	
	\textbf{Magnetization experiments.} Magnetization measurements were made on single-crystal samples down to $T=2$~K under applied magnetic fields up to $\mu_{0}H=5$~T using a Quantum Design, Inc.\ Superconducting Quantum Interference Device.   
	
	\textbf{Neutron diffraction experiments.} Neutron diffraction measurements were made on the TRIAX triple-axis spectrometer at the University of Missouri Research Reactor, and on the time-of-flight (TOF) CORELLI spectrometer at the Spallation Neutron Source at Oak Ridge National laboratory \cite{Ref_19}.  Single-crystal samples with masses of 4.5 and 11.6~mg were used on TRIAX and CORELLI, respectively. The samples were attached to Al sample mounts and cooled down to $T=6$~K using closed-cycle He refrigerators.  The integrated intensities of recorded Bragg peaks were corrected for the effects of neutron absorption using the \textsc{mag$2$pol}~\cite{Ref_27} software according to the procedure described in Ref.~\cite{Ref_28}.  The samples were flat rectangular plates with thicknesses of $0.1$~mm along $[001]$.  The TRIAX sample has an approximate width of $3$~mm along $[100]$ and approximate length of $2$~mm along $[010]$.  The CORELLI sample has an approximate width of $3$~mm along $[110]$ and approximate length of $3$~mm along $[1\bar{1}0]$.
	
	The TRIAX experiments were performed in elastic mode with incident and final neutron energies of $E_{\text{i},\text{f}}=14.7$ and $30.5$~meV. S\"{o}ller slit collimators with divergences of $60^{\prime}$-$60^{\prime}$-$80^{\prime}$-$80^{\prime}$ were inserted before the pyrolytic graphite (PG) $(002)$ monochromator, between the monochromator and sample, between the sample and PG $(002)$ analyzer, and between the analyzer and detector, respectively. PG filters were inserted before and after the sample to eliminate higher-order beam harmonics.  The sample's $(h0l)$ reciprocal-lattice plane was aligned within the horizontal scattering plane. In order to collect data for purely nuclear and magnetic Bragg peaks, a series of $(h0l)$ reflections were recorded at $T=30$ and $6$~K, i.e.\ significantly above $T_{\text{N}1}$ and below $T_{\text{N}2}$, respectively. Each Bragg peak was fitted with a Gaussian lineshape in order to determine its integrated intensity (area), center, and full width at half maximum. The magnetic signals of the low-temperature $\bm{\tau}_2$ Bragg peaks were then extracted by subtracting the integrated intensity of the $30$~K nuclear contribution when appropriate. A Lorentz factor ($L=\frac{1}{\sin{2\theta}}$) was also applied. Refinements were made with \textsc{fullprof}~\cite{Ref_29}. Throughout this work, error bars represent one standard deviation.

	 CORELLI experiments were performed at several temperatures above and below $T_{\text{N}1}$ and $T_{\text{N}2}$ with the $(h0l)$ plane set horizontal. However, the spectrometer employs $2$-D position-sensitive detectors which allow for recording some Bragg peaks above and below this plane.  The TOF measurement technique employs a white neutron beam, and the sample was rotated around its vertical axis at discrete intervals in order to cover a large swath of reciprocal space.  Individual Bragg peaks were obtained by performing cuts through the $2$-D maps recorded by the instrument, returning intensity versus $\mathbf{Q}$ profiles of the peaks. Integrated intensities were in turn calculated by adding up all of the data points located in the peak profile that were above background.  To account for the pulsed beam's non-linear distribution of wavelengths, a normalization to vanadium scattering was applied.  CORELLI data used for the structural and magnetic refinements presented in Supplementary Note~1 are restricted to a bandwidth of incoming neutron energies of $45$--$55$~meV in order to accurately correct for neutron absorption.

	\textbf{Electronic-band structure calculations.} Band structures using density functional theory with spin-orbit coupling were calculated with the Perdew-Burke-Ernzerhof (PBE) exchange-correlation functional, a plane-wave basis set, and the projected-augmented-wave method as implemented in VASP~\cite{Ref_30,Ref_31}. To account for the half-filled strongly-localized Eu $4f$ orbitals, a Hubbard-like~\cite{Ref_32} $U$ value of $5.0$~eV is used. For different helical magnetic structure with $\tau\approx(0,0,1/3)$, i.e.\ the hexagonal unit cell is tripled along the $\mathbf{c}$ axis with atoms fixed in their bulk positions. A Monkhorst-Pack $12\times12\times1$ $k$-point mesh with a Gaussian smearing of $0.05$~eV including the $\Gamma$ point and a kinetic-energy cutoff of $250$~eV have been used. To search for possible band gap closing points in the full Brillouin Zone, a tight-binding model based on maximally localized Wannier functions~\cite{Ref_33} was constructed to reproduce closely the bulk band structure including spin-orbit coupling in the range of $E_\text{F} \pm1$~eV with Eu $sdf$, In $sp$ and As $p$ orbitals as implemented in WannierTools~\cite{Ref_34}. 

\section{Data availability}
	The neutron scattering data that support our analysis of the ground state magnetic order in the EuIn$_2$As$_2$ are displayed in the Supplementary Note~1 and/or available in the MDF Open data repository with the identifiers DOI: 10.18126/u3j8-aplr and URL: https://doi.org/10.18126/u3j8-aplr.

	\section{Acknowledgements}
	This research was supported by the Center for Advancement of Topological Semimetals, an Energy Frontier Research Center funded by the U.S.\ Department of Energy Office of Science, Office of Basic Energy Sciences, through the Ames Laboratory under Contract No.\ DE-AC$02$-$07$CH$11358$. D.C.J., S.L.B., and A.K. were supported by U.S.\ Department of Energy Office of Science, Office of Basic Energy Sciences, Field Work Proposals at the Ames Laboratory operated under the same contract number.  This research used resources at the Spallation Neutron Source, a DOE Office of Science User Facility operated by the Oak Ridge National Laboratory.  This research used resources at the Missouri University Research Reactor.  Financial support for this work was provided by Fonds Qu\'{e}b\'{e}cois de la Recherche sur la Nature et les Technologies.  Much of this work was carried out while D.H.R.\ was on sabbatical at Iowa State University and Ames Laboratory, and their generous support under the U.S.\ Department of Energy Office of Science, Office of Basic Energy Sciences Contract No.\ DE-AC$02$-$07$CH$11358$ during this visit is gratefully acknowledged.
	
	\section{Author contributions}
	S.X.M.R,  B.G.U., R.J.M., T.W.H., and F.Y.\ conducted neutron scattering experiments.  S.X.M.R.\ refined the neutron diffraction data and determined the magnetic structure with input from A.K., B.G.U., D.C.J.,  and R.J.M.. B.K., S.L.B., and P.C.C. synthesized crystals and performed magnetization and resistance measurements. D.H.R.\ performed M\"{o}ssbauer measurements.  T.V.T., P.P.O., L.L.W, and A.V.\ performed symmetry analyses concerning the topologically protected properties.  L.L.W.\ performed electronic-band-structure calculations.

	\section{Additional information}
	Supplementary Information is available.
	
	\section{Competing interests}
	The authors declare no competing interests.

\clearpage
\newpage
\setcounter{equation}{0}
\setcounter{figure}{0}
\setcounter{table}{0}
\setcounter{page}{1}
\makeatletter

\maketitle

\begin{center}
\textbf{Magnetic crystalline-symmetry-protected axion electrodynamics and field-tunable unpinned Dirac cones in EuIn$_{\textbf2}$As$_\textbf{2}$} \\ \normalfont (Supplementary Information)
\end{center} 

\renewcommand{\figurename}{Supplementary Figure}
\renewcommand{\tablename}{Supplementary Table}
\renewcommand{\thesection}{Supplementary Note \arabic{section}}
%\tableofcontents
\section{S\lowercase{upplementary} N\lowercase{ote 1.}  Single-crystal neutron diffraction experiments}\label{Supp_Note_1}

In the following, we detail our determination of the $\mathbf{\textit{T}<\textit{T}_{\mathbf{N2}}}$ magnetic order using the TRIAX and CORELLI data.

\begin{figure}[h] %%% MSG Comparison%%
	\centering
	\includegraphics[width=0.67\textwidth,trim=0.0cm 0.0cm 0.0cm 0.0cm]{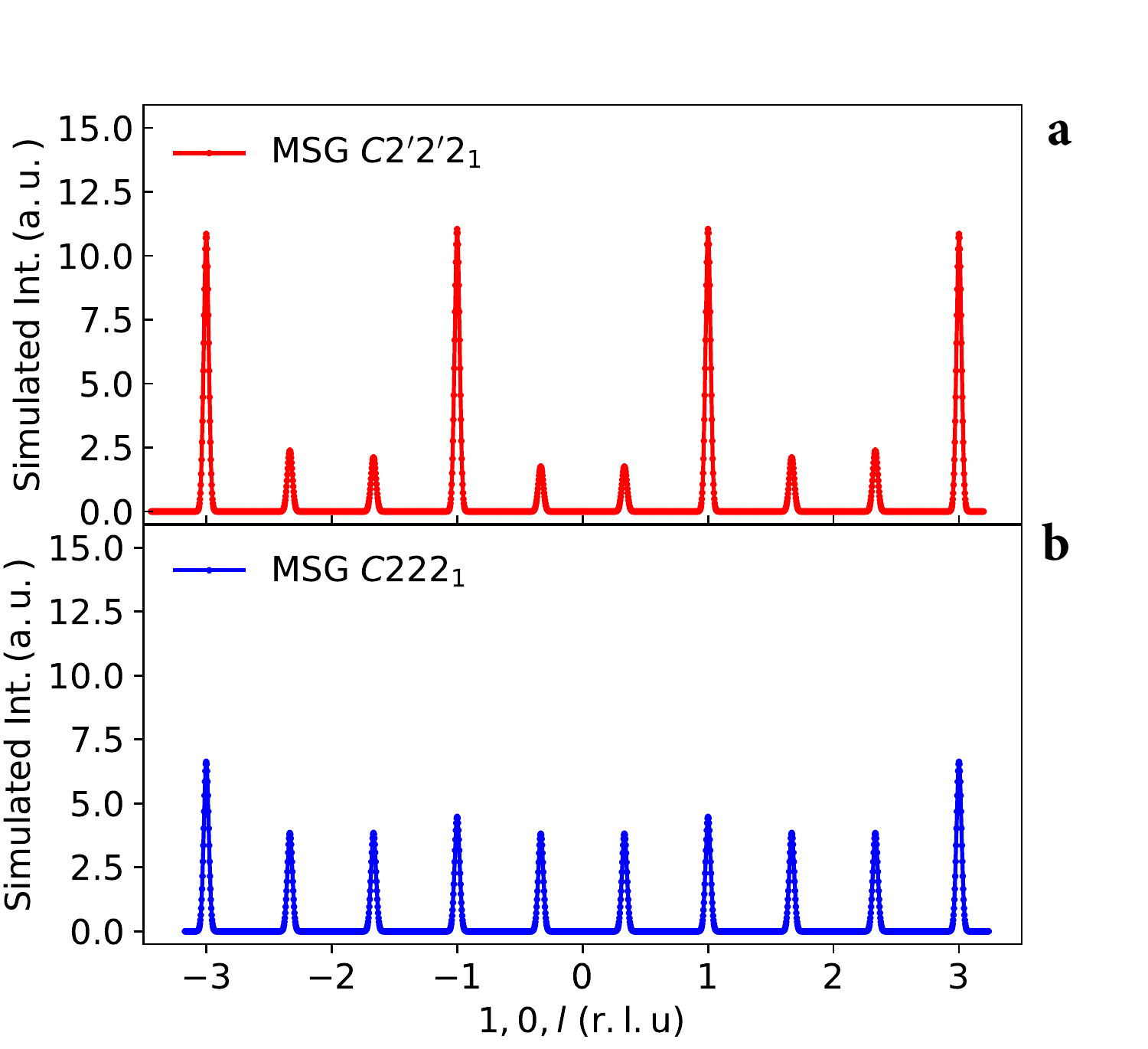}
	\caption{ \textbf{Simulated magnetic neutron diffraction patterns for broken-helix order with $\bm{\mu}\perp\mathbf{c}$. }\textbf{a,b}, Simulated pattern for magnetic space group (MSG) $C2^{\prime}2^{\prime}2_{1}$ (a) and MSG $C222_{1}$ (b) with helical-turn angles $\phi_{\text{rr}}=-80\degree$, $\phi_{\text{rb}}=130\degree$, and $\mu=6.0~\mu_{\text{B}}$.}
	\label{MSG_comparison}
\end{figure} 
 
\begin{figure}[h] %%% CORELLI and TRIAX data %%%
	\centering
	\includegraphics[width=0.67\textwidth,trim=0.0cm 0.0cm 0.0cm 0.0cm]{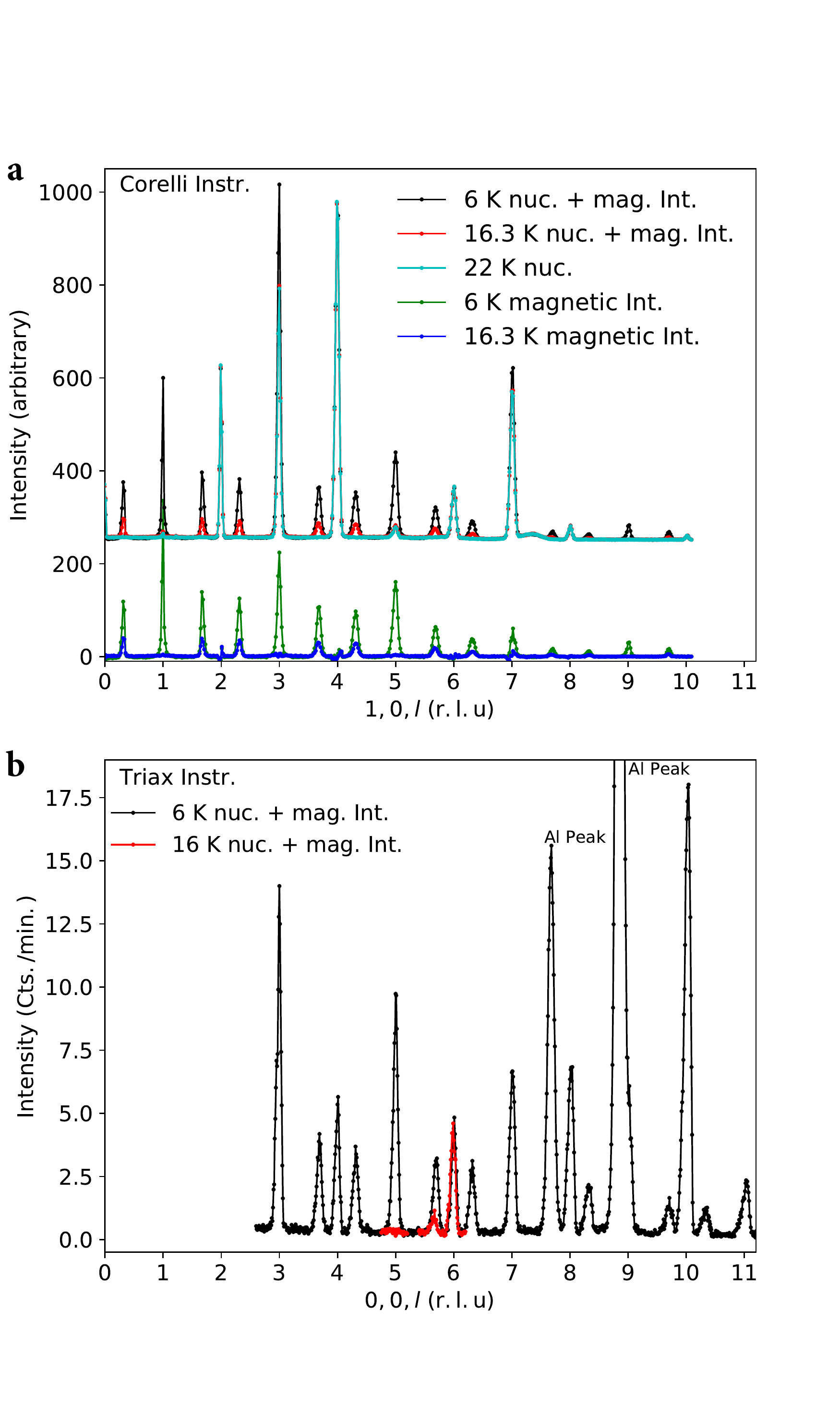}
	\caption {\textbf{Diffraction patterns from measurements made on CORELLI and TRIAX.} \textbf{a}, Line cuts of CORELLI data along $(10l)$ collected at $T=6$ (black), $16.3$ (red) and $22$~K (cyan). For clarity, a $250$~counts/minute offset has been applied. The $16.3$ (blue) and $6$~K (green) data not offset are the results of subtracting off the $22$~K data in order to visualize the magnetic diffraction. \textbf{b},  $(00l)$ line cut collected on TRIAX at $6$ (black) and $16$~K (red). }
	\label{line_cuts}
\end{figure} 
 
We first detail the symmetry analysis and magnetic space group (MSG) determination procedures which utilized the Bilbao Crystallography Server~\cite{Ref_1} and  simulations of diffraction patterns made with \textsc{mag$2$pol} \cite{Ref_2}.  Combinations of two distinct magnetic phases individually corresponding to $\bm{\tau}_{1}$ [approximated to $\bm{\tau_{1}}=(0,0,\frac{1}{3})$ ] and $\bm{\tau}_2=(0,0,1)$ were discarded as they returned spin-density-wave rather than local-moment type AF order.  Local-moment type AF order is more accurate than itinerant AF order based on our band-structure calculations indicating that bands with significant Eu $4f$ character lie well below the Fermi energy $E_{\text{F}}$.

\begin{table}
	\caption{\textbf{The Eu crystallographic sites in magnetic space group $\bm{C2^{\prime}2^{\prime}2_{1}}$ (No.\ $\bm{20.33}$).}  Positions are given in the setting of the parent space group $P$6$_{3}$/$mmc$ (No.\ $194$) with lattice parameters $(a,b,3c)$. $m_i$ are components of the ordered magnetic moment $\bm{\mu}$.  This table was created using the  Bilbao Crystallography Server~\cite{Ref_1}.}
	\label{Table_MSG_broken_helix}
	\resizebox{\textwidth}{!}{
		\begin{tabular}{|c|c|c|c|}
			\hline
			Atoms &  Coordinates with Magnetic Moments& Multiplicity &  \begin{tabular}[c]{@{}c@{}}$m_x$, $m_y$, $m_z$\\ ($\mu_{\text{B}}/\text{Eu}$)\end{tabular}\\
			\hline
			Eu(red) & \begin{tabular}[c]{@{}c@{}}$(0,0,0~\vert~m_x,m_y,m_z)  (0,0,\frac{1}{6}~\vert~m_y,m_x,m_z)$ \\ 
				$(0,0,\frac{1}{2}~\vert~-m_x,-m_y,m_z)  (0,0,\frac{2}{3}~\vert~-m_y,-m_x,m_z)$\end{tabular} & $4$ &  \begin{tabular}[c]{@{}c@{}}$m_x = 2.35$\\ 
				$m_y = 6.69$\\
				$m_z = 0.00$\end{tabular} \\
			\hline
			Eu(blue) & $(0,0,\frac{1}{3}~\vert~m_x,-m_x,m_z)  (0,0,\frac{5}{6}~\vert~-m_x,m_x,m_z)$ & $2$ & \begin{tabular}[c]{@{}c@{}}$m_x = -3.39$\\
				$m_z = 0.00$\end{tabular} \\
			\hline
		\end{tabular}
	}
\end{table}

We next searched for a single MSG with symmetry elements leading to reflection conditions consistent with both $\bm{\tau}_{1}$ and $\bm{\tau}_{2}$ that is also a subgroup of the paramagnetic grey MSG $P6_{3}/mmc1^{\prime}$ (No.\ $194.264$). We found that MSGs $Cmcm$ (No.\ $63.457$), $Cm^{\prime}c^{\prime}m$ (No.\ $63.452$), $C2^{\prime}2^{\prime}2_{1}$ (No.\ $20.33$), $C222_{1}$ (No.\ $20.31$), and $P2_1/m$ (No.\ $11.50$) are consistent with both propagation vectors. Note that additional MSGs are consistent with these propagation vectors but require further symmetry reductions and were thus disregarded. We ruled out $Cmcm$, $Cm^{\prime}c^{\prime}m$, $C222_{1}$ and $P2_1/m$ by simulating diffraction patterns for both $\bm{\mu}\perp\mathbf{c}$ and $\bm{\mu}\parallel\mathbf{c}$ and comparing them to $T=6$~K data.  Examples of simulated profiles are shown in  Supplementary Figure~\ref{MSG_comparison} which can be compared to $6$~K diffraction data in  Supplementary Figure~\ref{line_cuts}a.  In particular,  Supplementary Figure~\ref{MSG_comparison} illustrates the clear difference between the relative heights of commensurate and incommensurate magnetic Bragg peaks for $C2^{\prime}2^{\prime}2_{1}$ and $C222_{1}$, revealing the incompatibility of the latter with the $6$~K data in  Supplementary Figure~\ref{line_cuts}a. We are thus left with $C2^{\prime}2^{\prime}2_{1}$ as the MSG for the $T<T_{\text{N}2}$ AF order. Supplementary Table~\ref{Table_MSG_broken_helix} shows the positions for the two Eu sites [referred to as Eu(red) and Eu(blue)] in $C2^{\prime}2^{\prime}2_{1}$ assuming a tripling along $\mathbf{c}$ of the hexagonal chemical-unit cell.  Constraints imposed by the MSG on the orientation of the ordered magnetic moment associated with each Eu site are also indicated.

\begin{figure}[] %%% \chi, M, \rho%%%
	\centering
	\includegraphics[trim = 0mm 0mm 0mm 0mm,width=0.8\columnwidth]{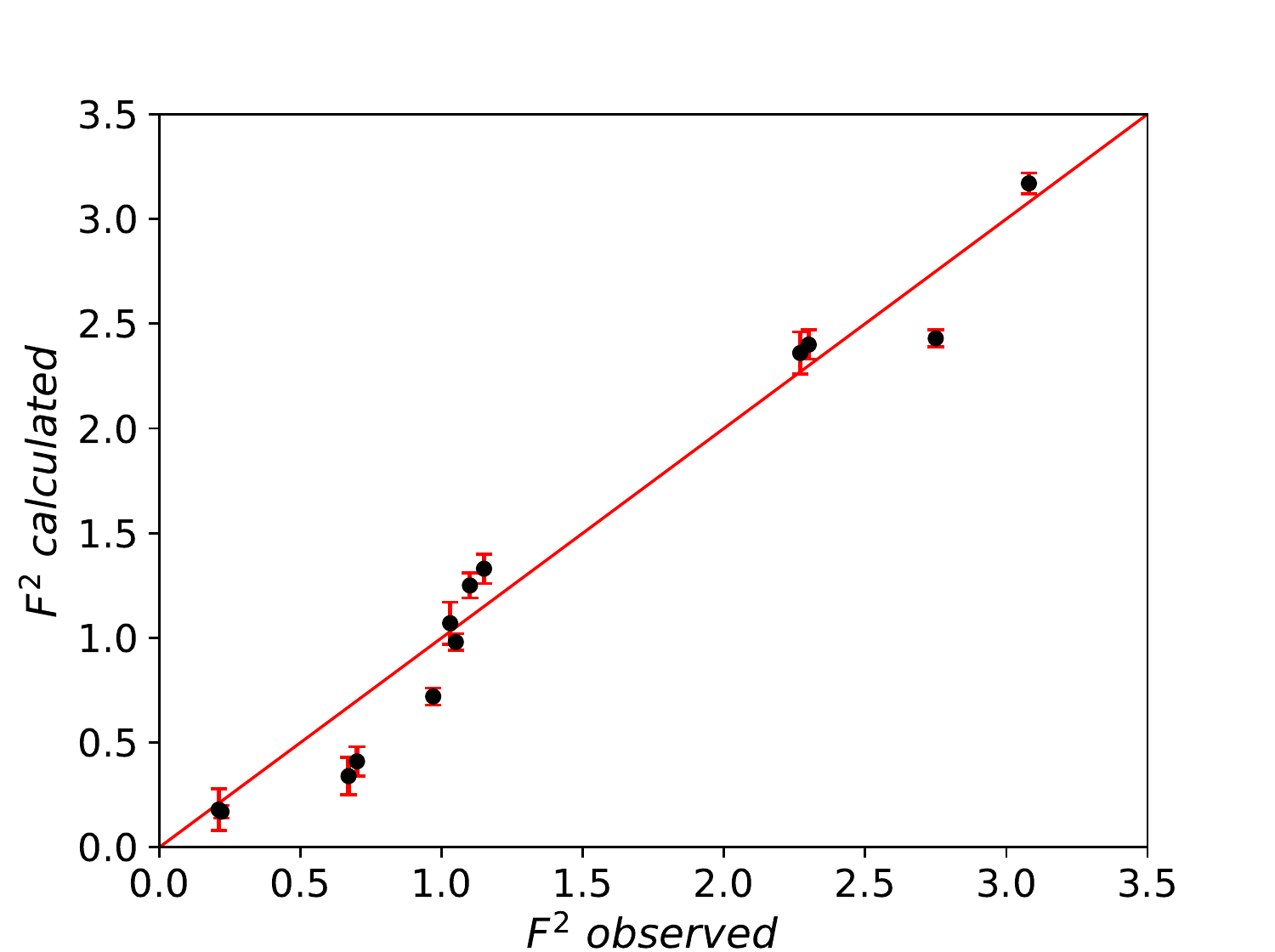}
	\caption[]{\textbf{Calculated versus observed square of the magnetic structure factor $\bm{\mathit{F}^2}$ for the refinement to data in Supplementary Table~\ref{TRIAX_data_table} using MSG $\bm{C2^{\prime}2^{\prime}2_{1}}$.} The line indicates the refinement, which has a goodness of fit of $R_{\text{F}}=7.5$. }
	\label{goodness_fit}
\end{figure}

\begin{figure}[] %%% Broken-helix order%%%
	\centering
	\includegraphics[trim=0cm 5.0cm 0.0cm 5.0cm, angle=90,origin=c]{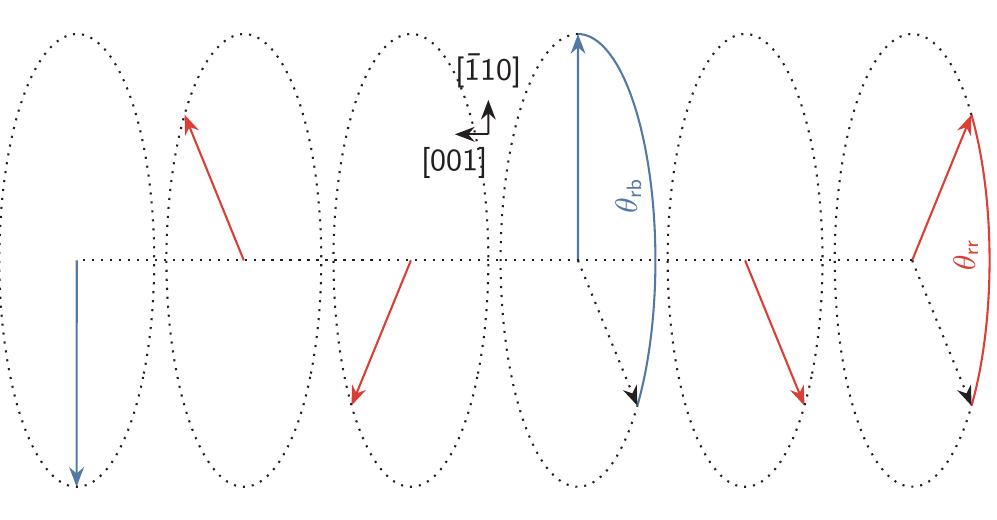}
	\caption {\textbf{Diagram of the broken-helix order.} The ordered moment lies $\perp[001]$, and the ferromagnetically aligned Eu layers are labeled red or blue, with the blue layers constrained to point along either $[\bar{1}10]$ or $-[\bar{1}10]$. $\phi_{\text{rr}}$ and $\phi_{\text{rb}}$ are the helical-turn angles between successive red-red and red-blue layers, respectively. }
	\label{bH_order}
\end{figure}

We next detail the steps of the refinements.   Examples of TRIAX data not corrected for neutron absorption are shown in Supplementary Figs.~\ref{line_cuts}b and \ref{TRIAX_data}.  Supplementary Tables~\ref{TRIAX_data_table_nuc} and \ref{TRIAX_data_table} list the Bragg peaks used for the refinements along with their integrated intensities and the corresponding absorption-corrected values.  First, a single-crystal refinement to the absorption corrected $T=30$~K data [corresponding to the paramagnetic (PM) phase] using the known chemical structure of the material~\cite{Ref_3,Ref_4} was performed allowing the scale factor and atomic positions to vary. The site occupations and thermal parameters could not be refined. The magnetic structure was then refined using the $6$~K absorption-corrected data corresponding to the magnetic order.  This refinement used the magnetic symmetry files created on the Bilbao crystallography server \cite{Ref_1} for $C2^{\prime}2^{\prime}2_{1}$ and the parameters found from the refinement to the $30$~K data. To account for the localized nature of the Eu$^{2+}$ magnetic moments, a refinement constraint of equal values of the total ordered magnetic moment $\mu$ for both sites was used.  Supplementary Figure~\ref{goodness_fit} illustrates the refinement which returned a goodness-of-fit value of $R_{\text{F}}=7.50$. The determined magnetic structure at $6$~K is the broken-helix order diagrammed in Fig.~\ref{Diff}d as well as Supplementary Figure~\ref{bH_order} with turn angles of $\phi_{\text{rr}}=-80(2)\degree$ and $\phi_{\text{rb}}=130(1)\degree$,  and $\mu=5.9(2)~\mu_{\text{B}}/\text{Eu}$ lying within the $\mathbf{ab}$ plane.

\begin{figure}[] %%% \chi, M, \rho%%%
	\centering
	\includegraphics[trim = 0mm 0mm 0mm 0mm,width=0.8\columnwidth]{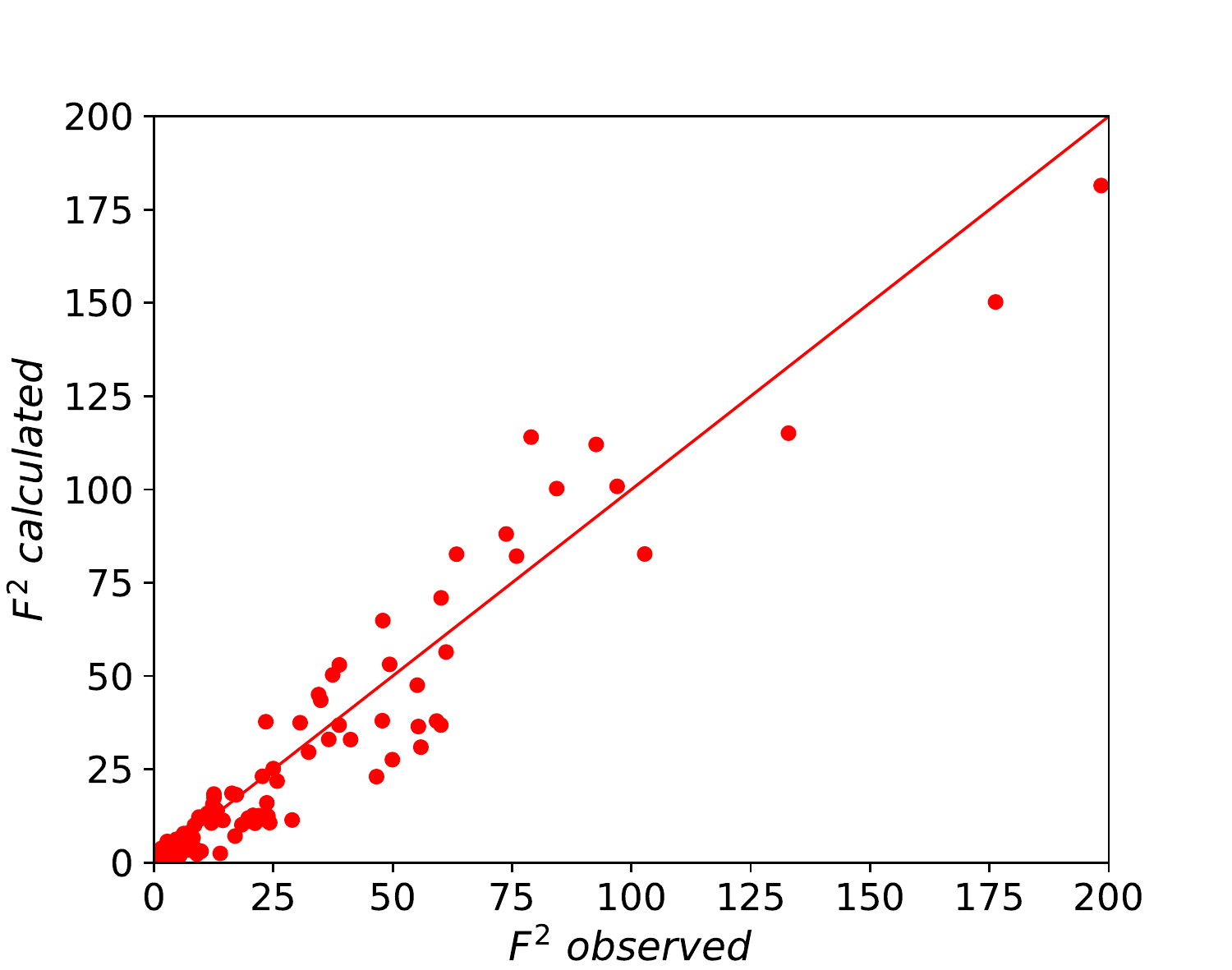}
	\caption[]{\textbf{Calculated versus observed square of the magnetic structure factor $\bm{\mathit{F}^2}$ for the refinement to $\bm{6}$~K CORELLI data using MSG $\bm{C2^{\prime}2^{\prime}2_{1}}$.} The line indicates the refinement, which has a goodness of fit of $R_{\text{F}}=17.2$. }
	\label{goodness_fit_corelli}
\end{figure}

This result is confirmed by a refinement to $T=6$~K data collected on the CORELLI instrument for a different sample.  Data corresponding to a reduced incident neutron energy bandwidth centered at $E=50$~meV ($1.54$~\AA) and spanning $45$--$55$~meV were used in order to accurately perform a correction for neutron absorption using \textsc{mag2pol} \cite{Ref_2}. The absorption correction and refinement procedures employed were similar to these previously described for the TRIAX data analysis. Sets of $18$ nuclear ($22$~K) and $108$ magnetic ($6$~K) independent reflections were used. We obtained turn angles of $\phi_{\text{rr}}=-68(2)\degree$ and $\phi_{\text{rb}}=124(1)\degree$, $\mu=6.1(2)~\mu_{\text{B}}/\text{Eu}$ and a goodness-of-fit value of $R_{\text{F}}=17.2$. Figure~\ref{goodness_fit_corelli} illustrates the quality of this refinement.  

\begin{figure}[] %%% TRIAX data examples%%%
	\centering
	\includegraphics[trim = 0mm 0mm 0mm 0mm,width=1.0\columnwidth]{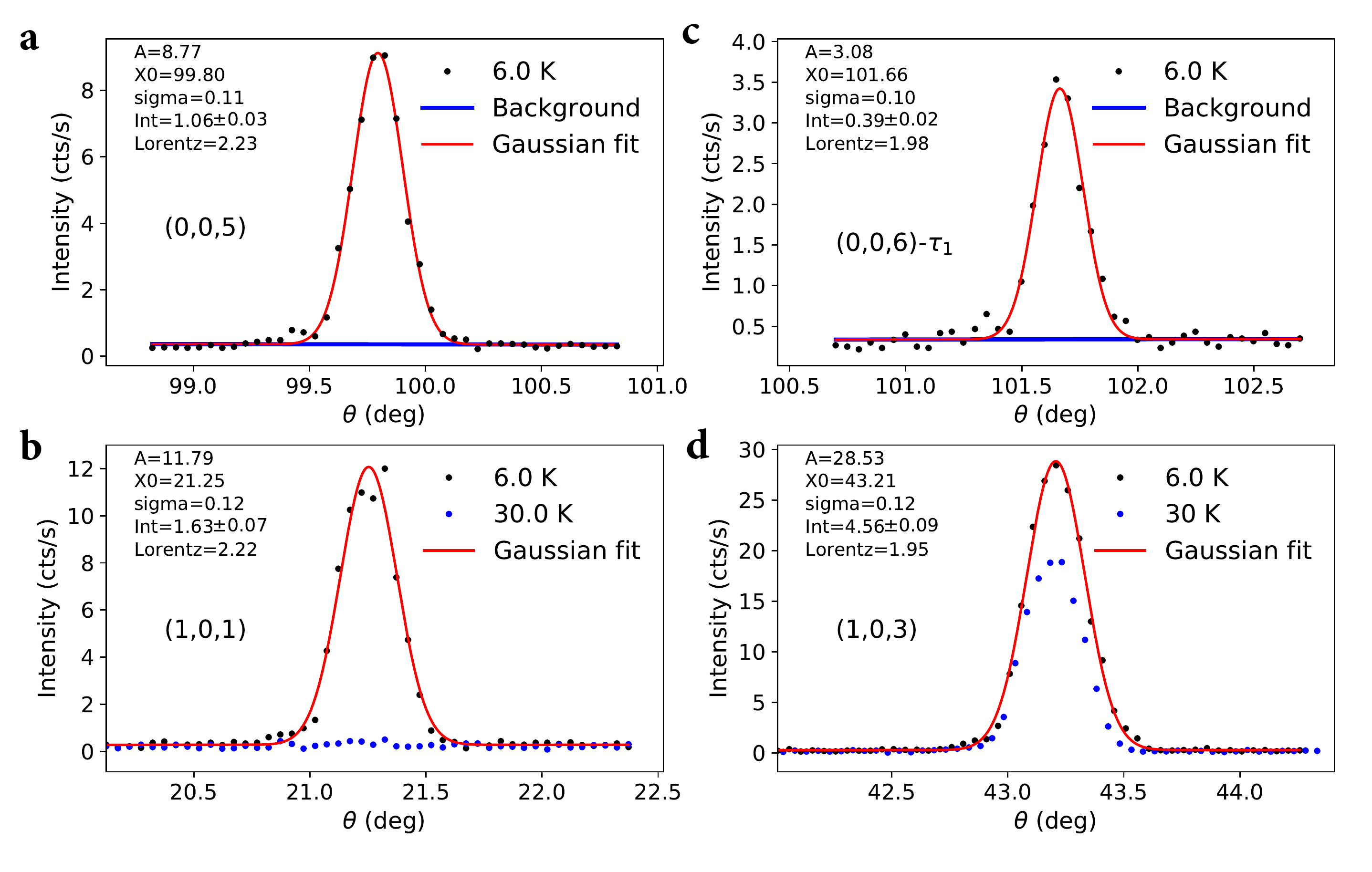}
	\caption[]{\textbf{Examples of Bragg peaks collected on TRIAX.} \textbf{a--d},  Measured profiles for $T=6$ and $30$~K. The $30$~K data are substituted by background lines for purely magnetic reflections in \textbf{a} and \textbf{c}. Each peak is fit using the combination of a Gaussian lineshape [y=$A \exp (-\frac{(x-x_0)^2)}{2 \sigma^2})$] and a linear background. The obtained fit parameters and the corresponding integrated intensities corrected by the Lorentz factor (labeled as Int) are displayed in each panel.}
	\label{TRIAX_data}
\end{figure}

\begin{table}
	\caption{\textbf{Integrated intensities for Bragg peaks recorded on TRIAX at $\bm{T=30}$~K.} The correction for neutron absorption was applied using \textsc{mag$2$pol} \cite{Ref_2} as described in Ref.~\cite{Ref_5} . }
	\label{TRIAX_data_table_nuc}
		\begin{tabular}{|c|c|c|}
			\hline
			$(h,0,l)$&  Integrated Intensity & Absorption Corr. Intensity \\
			\hline
			$(1,0,0)$ & $0.29 \pm 0.02$ & $0.41 \pm 0.03$ \\
			$(1,0,2)$ & $1.70 \pm 0.05$ & $2.47 \pm 0.07$ \\
			$(1,0,3)$ & $3.08 \pm 0.06$ & $4.65\pm 0.09$ \\
			$(0,0,6)$ & $0.60 \pm 0.03$ & $1.53 \pm 0.08$ \\
			$(1,0,4)$ & $5.04 \pm 0.21$ & $8.08 \pm 0.34$ \\
			$(2,0,0)$ & $0.23 \pm 0.02$ & $0.33 \pm 0.03$ \\
			$(2,0,2)$ & $1.22 \pm 0.03$ & $1.79 \pm 0.04$ \\
			$(2,0,3)$ & $1.96 \pm 0.03$ & $2.93 \pm 0.04$ \\
			$(2,0,4)$ & $2.72 \pm 0.10$ & $4.17 \pm 0.15$ \\
			\hline
		\end{tabular}

\end{table}

\begin{table}
	\caption{\textbf{Integrated intensities for Bragg peaks recorded on TRIAX at $\bm{T=6}$~K.} The correction for neutron absorption was applied using \textsc{mag$2$pol} \cite{Ref_2} as described in Ref.~\cite{Ref_5}. When present, nuclear contributions to the Bragg peaks were subtracted using the $30$~K data given in Table.~\ref{TRIAX_data_table_nuc}. }
	\label{TRIAX_data_table}
	\begin{tabular}{|c|c|c|}
		\hline
		$(h,0,l)$&  Integrated Intensity & Absorption Corr. Intensity\\
		\hline
		$(0,0,4)-\bm{\tau}_{1}$ & $0.31 \pm 0.02$  &$1.15\pm 0.07$ \\
		$(0,0,4)+\bm{\tau}_{1}$ & $0.34 \pm 0.02$  &$1.10\pm 0.06$ \\
		$(1,0,0)+\bm{\tau}_{1}$ & $0.48 \pm 0.03$  &$0.67\pm 0.04$ \\
		$(0,0,5)$ & $1.06 \pm 0.03$  & $3.08 \pm 0.09$ \\
		$(1,0,1)$ & $1.63 \pm 0.07$  &$2.30 \pm 0.10$ \\
		$(1,0,2)-\bm{\tau}_{1}$ & $0.49 \pm 0.03$ &  $0.70 \pm 0.04$ \\
		$(1,0,3)$ & $1.50 \pm 0.07$ &  $2.27 \pm 0.10$ \\
		$(0,0,6)-\bm{\tau}_{1}$ & $0.39 \pm 0.02$ &  $1.03 \pm 0.05$ \\
		$(0,0,6)+\bm{\tau}_{1}$ & $0.43 \pm 0.03$ &  $1.05 \pm 0.07$ \\
		$(0,0,7)$ & $1.20 \pm 0.03$ &  $2.75 \pm 0.07$ \\
		$(0,0,8)+\bm{\tau}_{1}$	& $0.47 \pm 0.05$ &   $0.97 \pm 0.10$ \\
		$(2,0,2)+\bm{\tau}_{1}$	& $0.15 \pm 0.03$ &   $0.22 \pm 0.04$ \\
		$(2,0,4)-\bm{\tau}_{1}$	& $0.14 \pm 0.02$ &   $0.21 \pm 0.03$ \\
		\hline
	\end{tabular}
	
\end{table}

%\subsection{CORELLI investigation of the magnetic phases and their temperature dependencies}

In the following, we detail our determination of the $\mathbf{\textit{T}}_{\mathbf{N2}}\mathbf{<}\mathbf{\textit{T}}\mathbf{\leq}\mathbf{\textit{T}}_{\mathbf{N1}}$ magnetic order using TRIAX and CORELLI data.

A reliable refinement of the TRIAX data for $T_{\text{N}2}<T \leq T_{\text{N}1}$ could not be completed.  This is due to the rather weak magnetic intensities at these temperatures because of the smaller value of $\mu$ and the strong neutron absorption. Nevertheless, data in Fig.~\ref{fig_bloc}b and  Supplementary Figure~\ref{line_cuts} reveal the systematic absence of Bragg peaks corresponding to $\bm{\tau}_2$ for $T>T_{\text{N}2}$.  The presence of magnetic Bragg peaks at positions corresponding to $\bm{\tau}_1$ is in agreement with the stabilization of pure $60\degree$-helix order and the reflection conditions for MSGs $P6_12^{\prime}2^{\prime}$ (No.\ $178.159$), $P6_{1}22$ (No.\ $178.155$), $P6_52^{\prime}2^{\prime}$ (No.\ $179.165$), and $P6_{5}22$ (No.\ $179.161$). As shown by the group-subgroup graph given in  Supplementary Figure~\ref{MSG_tree} and discussed further below, MSGs $P6_{1}22$ and $P6_{5}22$ are inconsistent with the stabilization of broken-helix magnetic order with MSG $C2^{\prime}2^{\prime}2_1$ for $T<T_{\text{N}2}$. The remaining two MSGs ($P6_12^{\prime}2^{\prime}$ and $P6_52^{\prime}2^{\prime}$) represent right-handed and left-handed chiral versions of the same pure $60\degree$-helix order and cannot be differentiated with our data.  Using either one of these MSGs and the intensity of the $(0,0,6)-\bm{\tau}_1$ reflection measured on TRIAX at $16$~K (see  Supplementary Figure~\ref{line_cuts}b), we calculate that $\mu=2.0(2)~\mu_{\text{B}}/$Eu at $16$~K. 

\begin{figure}[] %%% \chi, M, \rho%%%
	\centering
	\includegraphics[trim = 0mm 0mm 0mm 0mm,width=0.8\columnwidth]{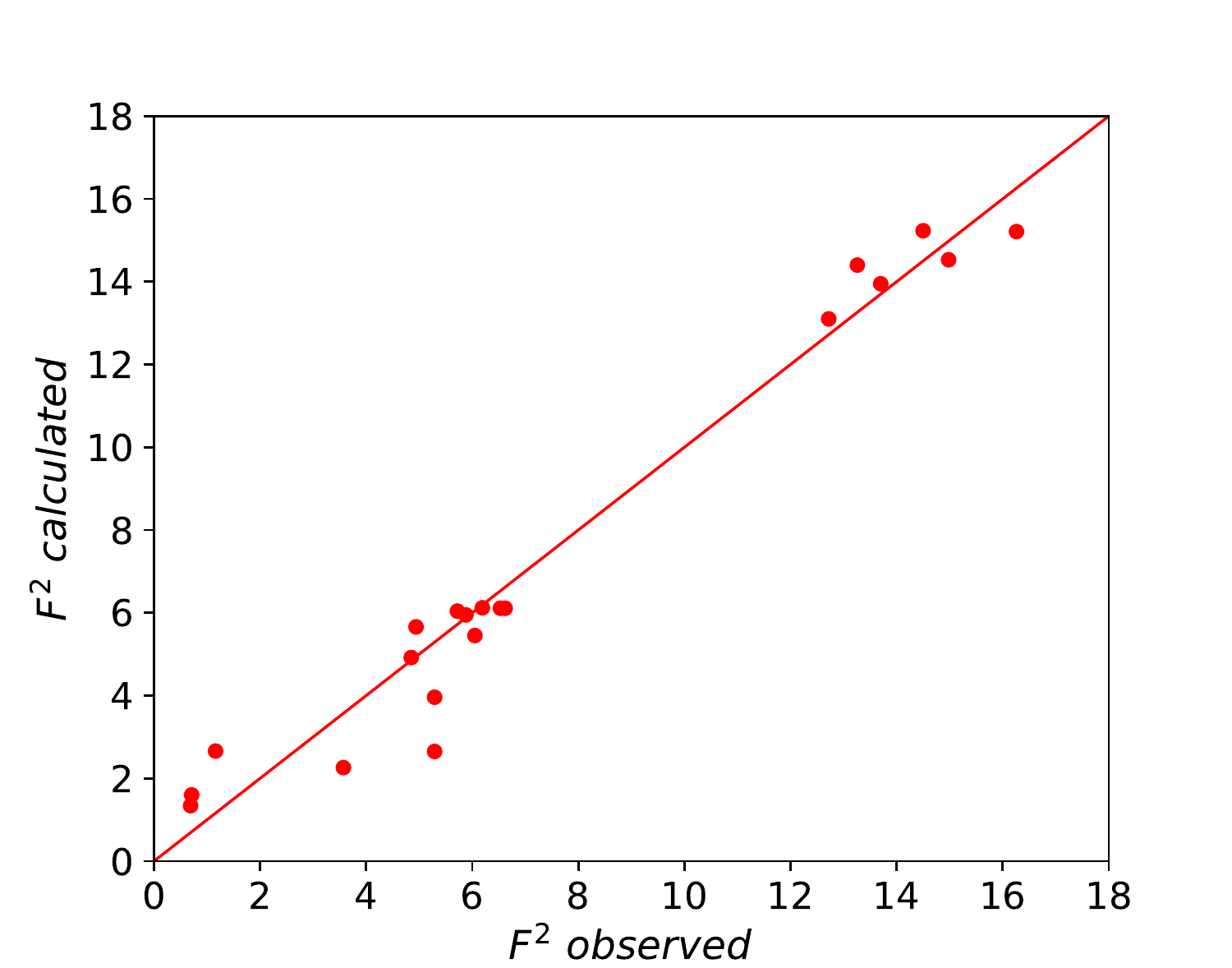}
	\caption[]{\textbf{Calculated versus observed squares of the magnetic structure factor $\bm{\mathit{F}^2}$ for the refinement to 16.3~K CORELLI data using MSG $P6_12^{\prime}2^{\prime}$.} The line indicates the refinement, which has a goodness of fit of $R_{\text{F}}=7.37$. }
	\label{goodness_fit_corelli_16pt3}
\end{figure}

To complement these results, we made a refinement for the $T_{\text{N}2}<T \leq T_{\text{N}1}$ magnetic phase using MSG $P6_12^{\prime}2^{\prime}$ and data for $20$ independent Bragg peaks corresponding to $\bm{\tau}_1$ taken with CORELLI at $16.3$~K.  Similar to above,  data corresponding to a reduced incident neutron energy bandwidth centered at $E=50$~meV ($1.54$~\AA) and spanning $45$--$55$~meV were used in order to accurately perform a correction for neutron absorption using \textsc{mag2pol} \cite{Ref_2}. In agreement with our TRIAX estimation, the refinement returns $\mu=2.5(5)~\mu_{\text{B}}/\text{Eu}$ with a goodness-of-fit value of $R_{\text{F}}=7.37$. Supplementary Figure~\ref{goodness_fit_corelli_16pt3} illustrates the quality of this refinement.

In the following, we detail our determination of the magnetic phases' temperature evolution using CORELLI data.

\begin{figure}[] %%% (1,0,0) Temperature Dependence%%%
	\centering
	\includegraphics[width=0.67\textwidth,trim=0.0cm 0.0cm 0.0cm 0.0cm]{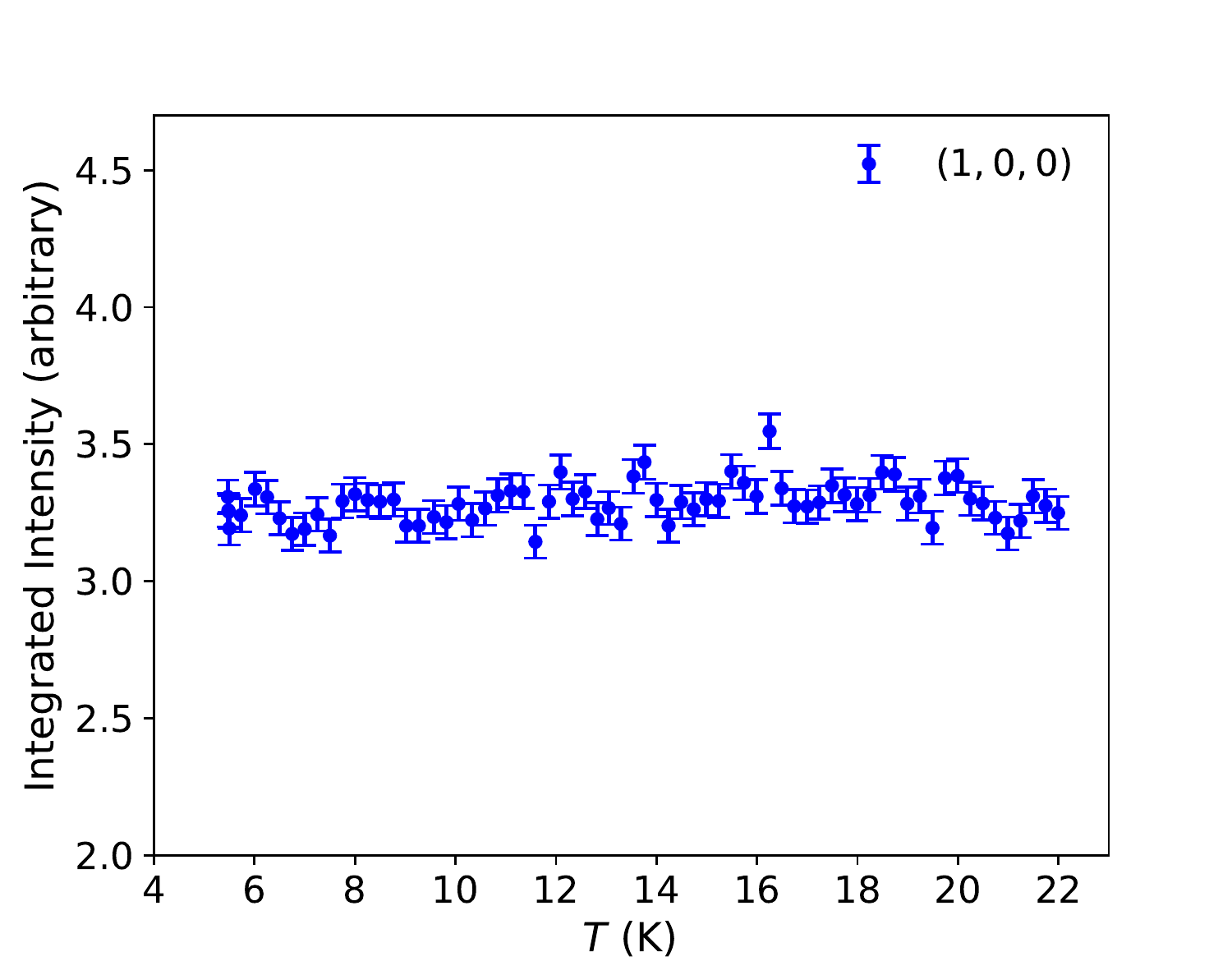}
	\caption {\textbf{Temperature dependence of the integrated intensity of the $\bm{(1,0,0)}$ Bragg peak measured on CORELLI .} The featureless temperature dependence between $T=5.5$ and $22$~K indicates an absence of a ferromagnetic moment pointing along $\mathbf{c}$.     }
	\label{100_order_para}
\end{figure}

The CORELLI data also allowed us to precisely and efficiently visualize the appearance of magnetic Bragg peaks and their positions upon cooling down to $T=6$~K. Figure~\ref{fig_bloc}b displays data from magnetic order parameter measurements and Fig.~\ref{fig_bloc}c displays the temperature evolution of the incommensurability associated with $\bm{\tau}_{1}$ below $T_{\text{N}1}$.  Large sets of data collected at $6$, $16.3$ and $22$~K are shown in Fig.~\ref{Diff}a and  Supplementary Figure~\ref{line_cuts}a. The positions of the magnetic peaks at $6$~K lead to the precise determination of the two antiferromagnetic (AF) propagation vectors: ($1$) $\bm{\tau}_{1}=(0,0,\tau_{1z})$ with $\tau_{1z}=0.303(1)$ corresponding to the incommensurate peaks; ($2$) $\bm{\tau}_{2}=(0,0,1)$ \cite{Ref_6} corresponding to commensurate magnetic Bragg peaks.  Note that for $\bm{\tau}_{2}$ some nuclear and magnetic Bragg peaks overlap. CORELLI data were also used to determine symmetry and ordered-moment-direction aspects of the magnetic phases. For instance, we observed the absence of magnetic Bragg peaks along the $(h00)$ direction.  Supplementary Figure~\ref{100_order_para} further shows that the $(1,0,0)$ Bragg peak has no significant temperature dependence which, based on our simulations, indicates that $\bm{\mu}$ does not have a ferromagnetic component along $\mathbf{c}$.  This information, along with the magnetization data shown in Fig.~\ref{Field_Pol}b and Supplementary Figure~\ref{characterization_2}b support the conclusion that $\bm{\mu}$ lies in the $\mathbf{ab}$ plane below $T_{\text{N}1}$.

%\subsection{TRIAX order parameter measurement}

\begin{figure}[] %%% \chi, M, \rho%%%
	\centering
	\includegraphics[trim = 0mm 0mm 0mm 0mm,width=0.8\columnwidth]{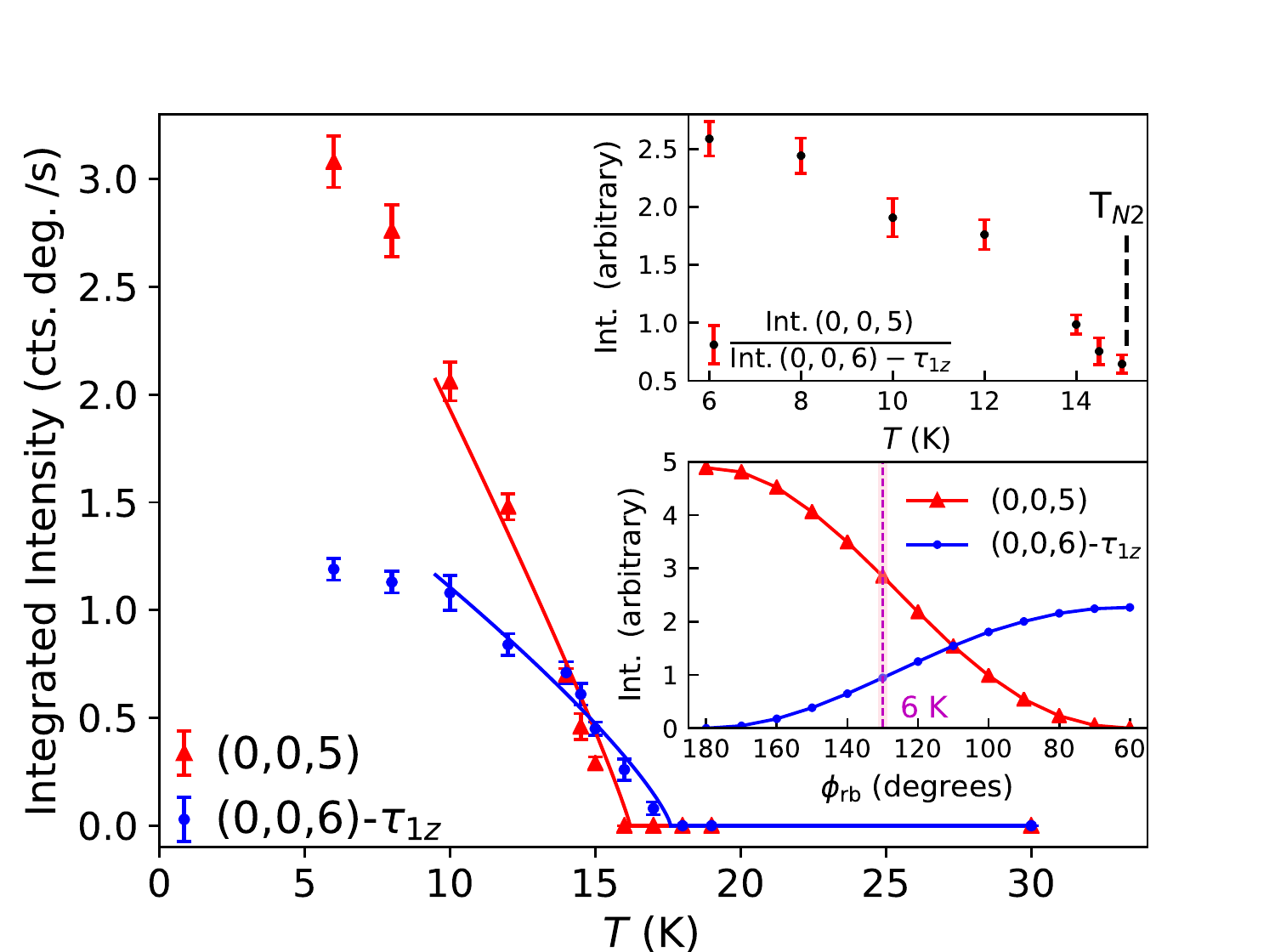}
	\caption[]{\textbf{Magnetic order parameters measured on TRIAX.} The temperature evolutions of the integrated intensities of the $(0,0,5)$ and $(0,0,6-\tau_{1z})$ Bragg peaks.  Lines are guides to the eye.  The top right inset displays the ratio of the integrated intensities of the two peaks. The bottom right inset shows a simulation of the peaks intensities as a function of the helix turn angle $\phi_{\text{rb}}$.}
	\label{TRIAX_order_para}
\end{figure}

Lastly, for comparison with Fig.~\ref{fig_bloc}b, absorption corrected TRIAX data for the temperature dependencies of the  $(0,0,5)$ and $(0,0,6-\bm{\tau}_{1z})$ Bragg peaks are displayed in  Supplementary Figure~\ref{TRIAX_order_para}.

\section{S\lowercase{upplementary} N\lowercase{ote 2.}  Magnetic susceptibility, magnetization, and resistance}

\begin{figure}[] %%% \chi, M, \rho Mossbauer%%%%
\centering
\includegraphics[trim = 0mm 12mm 0mm 28mm,width=0.5\columnwidth]{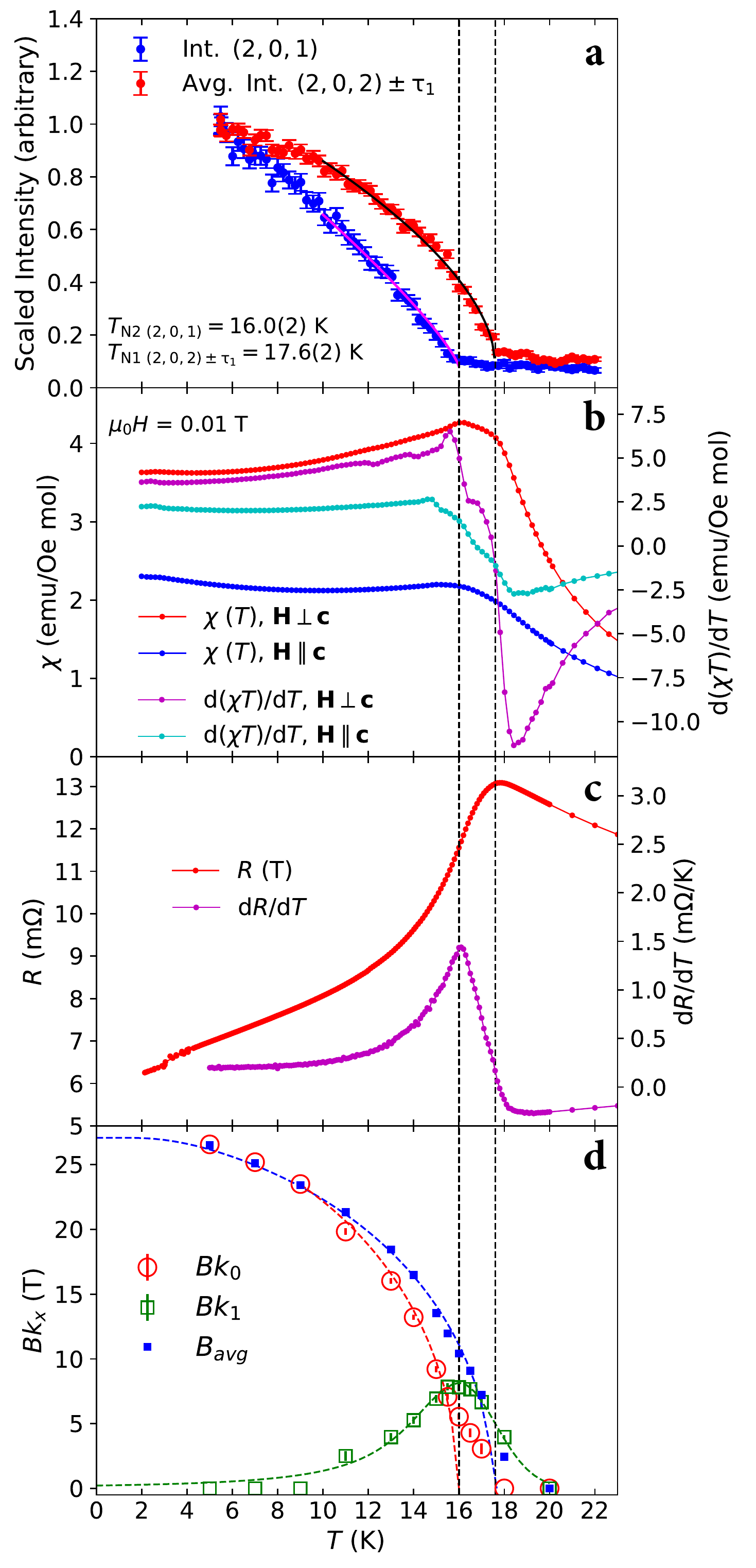}
\caption[]{\textbf{Temperature dependencies of magnetic order parameter, magnetic susceptibility, resistance, and \bm{$^{151}$}Eu hyperfine field.} \textbf{a}, Magnetic order parameter of the $(2,0,1)$ and $(2,0,2)\pm \tau_1$ Bragg peaks measured on CORELLI \cite{Ref_7} scaled to be $1$ at $T=6$~K. The N\'{e}el temperatures $T_{\text{N}1}$ and $T_{\text{N}2}$ found by neutron diffraction are indicated by vertical dashed lines through the different panels. \textbf{b}, Magnetic susceptibility (red+blue) and its derivative with respect to temperature (magenta+cyan). \textbf{c}, Temperature evolution of the electrical resistance (red), and its derivative with respect to temperature (magenta). \textbf{d}, Temperature dependence of the Fourier components of the hyperfine field ($B_{\text{hf}}$) from analysis of $^{151}$Eu M\"ossbauer spectra.}
\label{characterization_1}
\end{figure}

 Supplementary Figure~\ref{characterization_1}a reproduces the magnetic order parameter plots determined from neutron diffraction that are also shown in Fig.~\ref{fig_bloc}b.   We find features corresponding to the two N\'{e}el temperatures [$T_{\text{N}1}=17.6(2)$ and  $T_{\text{N}2}=16.2(1)$~K] in magnetization, resistance, and $^{151}$Eu M\"{o}ssbauer spectroscopy measurments as well. Supplementary Figure~\ref{characterization_1}b. shows the temperature dependence of the magnetic susceptibility $\chi=\frac{M}{H}$ and $\frac{\diff(\chi T)}{\diff T}$  for an applied magnetic field of $\mu_{0}H=0.01$~T. Changes in the slope of $\chi(T)$ are evident near the dashed lines indicating $T_{\text{N}1}$ and $T_{\text{N}2}$.

Standard $4$-wire resistance ($R$) measurements were made on single-crystal samples down to $T=2$~K with a Quantum Design, Inc.\ Physical Property Measurement System using Pt leads attached with Epotek H$20$E silver epoxy. Resistance data plotted in  Supplementary Figure~\ref{characterization_1}c show a loss of spin-disorder scattering upon cooling through $T_{\text{N}1}$ and a distinct change in slope upon further cooling through $T_{\text{N}2}$.

Next, Supplementary Fig.~\ref{characterization_1}d shows that the Fourier components of the hyperfine field found by $^{151}$Eu M\"{o}ssbauer spectroscopy measurements also show evidence for the two magnetic ordering transitions.  More details concerning the M\"{o}ssbauer measurements and the plots in Supplementary Figure~\ref{characterization_1}d are given in a separate section below.

\begin{figure}[] %%% CW and M(H) %%%
\centering
\includegraphics[trim = 0mm 0mm 0mm 0mm,width=0.5\columnwidth]{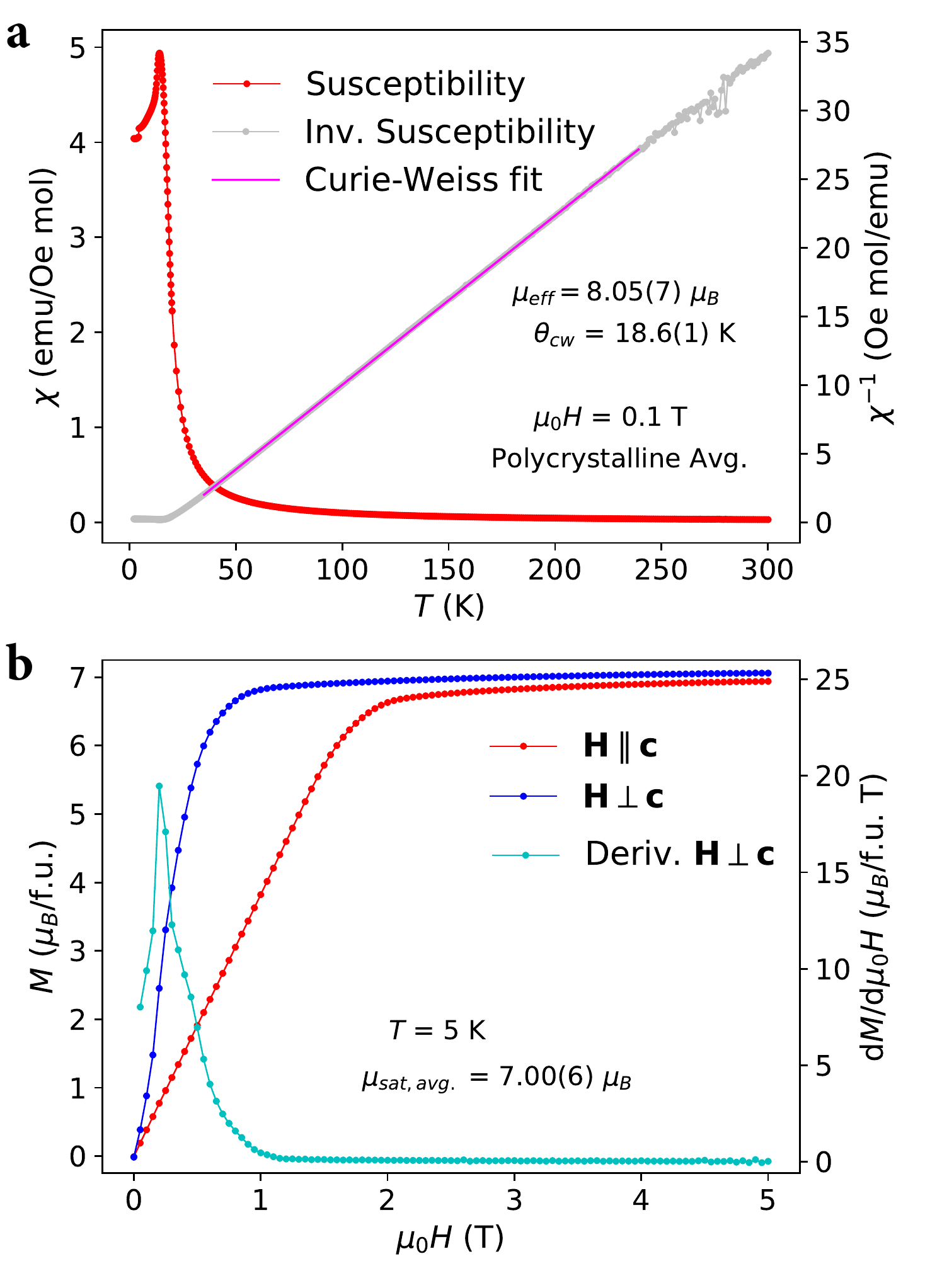}
\caption[]{\textbf{Curie-Weiss fit and magnetization versus magnetic field data at $\bm{T=5}$~K.} \textbf{a}, The magnetic susceptibility (red) and its inverse (grey) plotted versus temperature for an applied magnetic field of $\mu_{0}H=0.1$~T. The linear region of the inverse susceptibility is fitted with a Curie-Weiss model (fuchsia). \textbf{b}, Data from magnetization measurements performed for $\mathbf{H} \parallel \mathbf{c}$ (red) and $\mathbf{H} \perp \mathbf{c}$ (blue).  The field derivative of the $\mathbf{H} \perp \mathbf{c}$ data ($\frac{\diff M}{\diff H}$) is also shown (cyan). Magnetic anisotropy consistent with the ordered magnetic moment lying perpendicular to $\mathbf{c}$ is evident, and a saturation magnetization of $7.00(6)~ \mu_{\text{B}}/$f.u. is obtained. The $\mathbf{H} \perp \mathbf{c}$ data show an inflection point at $\mu_{0}H\approx0.2$~T that is visible as a  peak in $\frac{\diff M}{\diff H}$. }
\label{characterization_2}
\end{figure}

A Curie-Weiss fit to high-temperature $\chi(T)$ data is shown in  Supplementary Figure~\ref{characterization_2}a and yields an effective magnetic moment of $\mu_{\text{eff}}=8.05(7)~\mu_{\text{B}}/\text{Eu}$ and a Weiss temperature of $\theta_{\text{CW}}=18.6(1)$~K.  $\mu_{\text{eff}}$ is consistent with Eu$^{2+}$, whereas the positive value of $\theta_{\text{CW}}$ indicates predominantly ferromagnetic interactions and likely reflects the magnetic interactions associated with the ferromagnetically-aligned moments within the Eu planes.  The Curie-Weiss fits were performed after taking the appropriate polycrystalline  average of data  taken with $\mathbf{H}$ applied either perpendicular or parallel to the crystalline $\mathbf{c}$ axis.

Magnetization versus field measurements made with $\mathbf{H}\perp\mathbf{c}$  or $\mathbf{H}\parallel\mathbf{c}$ reveal  magnetic anisotropy consistent with magnetic moments ordered solely within the $\textbf{ab}$-plane.  These data appear in Fig.~\ref{Field_Pol}b and in Supplementary Figure~\ref{characterization_2}b. A saturated moment of $\mu_{\text{sat}}=7.00(6)~\mu_{\text{B}}/$f.u.\ is  achieved as expected for Eu$^{2+}$, where $M_{sat}=gS=7~\mu_{\text{B}}/$f.u.\ with $g=2$ and $S=7/2$.  $g$ is the spectroscopic splitting factor and $S$ is the spin.  Data for  $\mathbf{H}\perp\mathbf{c}$ show an inflection point at $H\approx2$~kOe which is seen as a sharp peak in $\frac{\diff M}{\diff H}$. This feature may indicate a continuous crossover from helical- to fan-type AF order with increasing field~\cite{Ref_8}.

\section{S\lowercase{upplementary} N\lowercase{ote 3.}  Details of the $^{\mathbf{151}}$E\lowercase{u} M\"ossbauer study }
%\subsection{Method}
	\begin{figure}[] %%% Mossbauer_spectra%%%
		\centering
		\includegraphics[width=0.7\textwidth,trim=0.0cm 0.0cm 0.0cm 0.0cm]{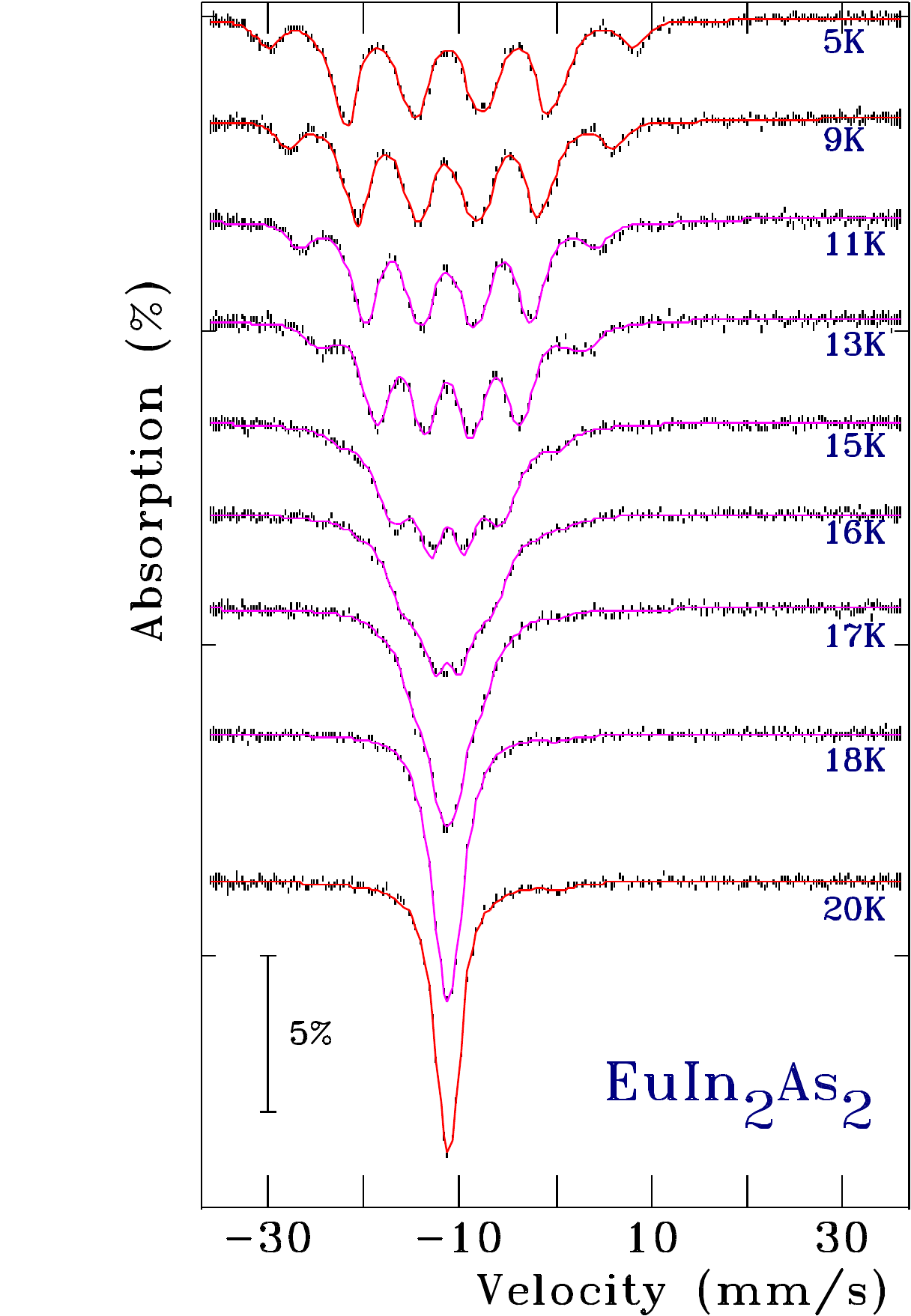}
		\caption {\textbf{$^{\bm{151}}$Eu M\"ossbauer spectra showing the evolution of the magnetic order on heating from a temperature of $\bm{T=5}$ to $\bm{20}$~K.} Solid lines are fits using the full Hamiltonian (red lines, $5$~K, $9$~K and $20$~K) or the incommensurate modulated model (magenta lines). A description of the fitting procedure is given in the text. }
		\label{Mossbauer_spectra}
	\end{figure}
	
$^{151}$Eu M\"ossbauer spectroscopy measurements were carried out using a $4$~GBq $^{151}$SmF$_3$ source driven in sine mode and calibrated using a standard $^{57}$Co\underline{Rh}/$\alpha$-Fe foil.  Isomer shifts are quoted relative to EuF$_3$ at ambient temperature. The sample was cooled in a vibration-isolated closed-cycle helium refrigerator with the sample in helium exchange gas. Spectra for different temperatures are shown in  Supplementary Figure~\ref{Mossbauer_spectra}.

In order to determine the hyperfine field $B_{\mathrm{hf}}$, the spectra in  Supplementary Figure~\ref{Mossbauer_spectra} are fit as follows. Spectra for $T\alt10$~K and $T\agt18$~K are fit to a sum of Lorentzian lineshapes with the positions and intensities derived from a full solution to the nuclear Hamiltonian \cite{Ref_9}. However,  spectra taken between $10$~K and $18$~K are fit using a model that derives a distribution of hyperfine fields from an (assumed) incommensurate magnetic structure with a sinusoidally modulated value for the ordered magnetic moment $\mu$ \cite{Ref_10, Ref_11}.  We next describe this distribution model. 

If we denote the antiferromagnetic (AF) propagation vector as $\mathbf{k}$ (instead of $\bm{\tau}$), assume that the modulation in $\mu$ along the direction of the propagation vector can be written in terms of its Fourier components, and that the  hyperfine field is a linear function of $\mu$ at any given site, then the variation of $B_{\text{hf}}$ with distance $x$ along $\mathbf{k}$ can be written as:\cite{Ref_10}
\begin{equation}
B_{\text{hf}}(kx) = Bk_0 + \sum^n_{l=0} Bk_{2l+1} \sin[(2l+1)kx]\ . \label{eqn:fourier}
\end{equation}
$Bk_{n}$ are the odd Fourier coefficients of the field modulation and $kx$ is a position in reciprocal space along the direction of $\mathbf{k}$. As $+B_{\text{hf}}$ and $-B_{\text{hf}}$ are indistinguishable, $kx$ only needs to run over half the modulation period. Variations of this modeling have been used to fit spectra for EuPdSb \cite{Ref_10} and Eu$_4$PdMg \cite{Ref_12}.

	\begin{figure}[] %%% Mossbauer_Bhf_vs_kx%%%
		\centering
		\includegraphics[width=0.7\textwidth,trim=0.0cm 0.0cm 0.0cm 0.0cm]{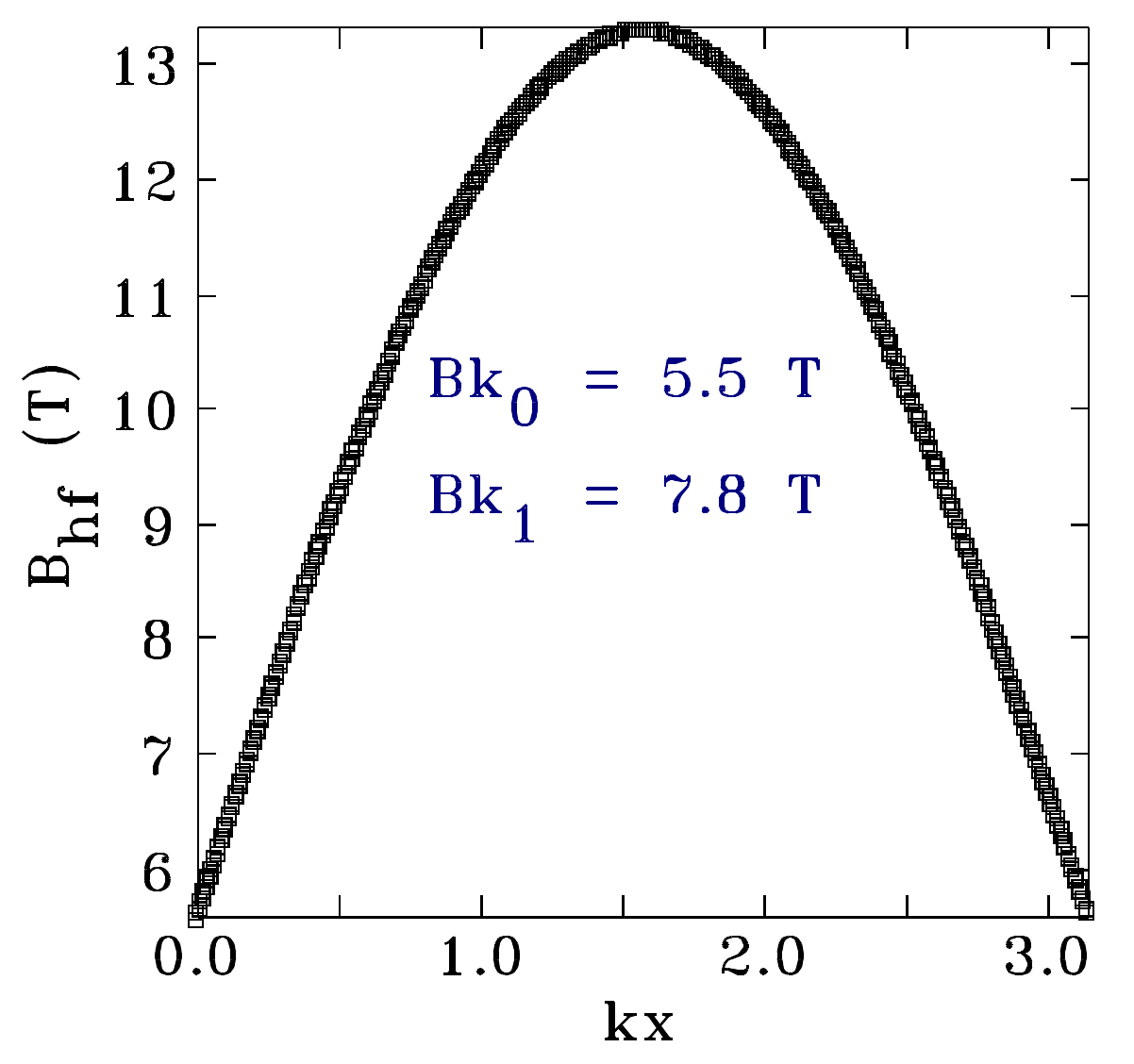}
		\caption {\textbf{Magnitude of the hyperfine magnetic field as a function of distance along the direction of the magnetic-propagation vector for $\bm{T=16}$~K.} $kx=\pi$ corresponds to the length of the Brillouin zone along $\bm{kx}$. Values of the Fourier components $Bk_0$ and $Bk_1$ discussed in the text are indicated.}
		\label{Mossbauer_Bhf_kx}
			\end{figure}

	\begin{figure}[] %%% Mossbauer_Fourier_Histograms%%%
		\centering
		\includegraphics[width=0.7\textwidth,trim=0.0cm 0.0cm 0.0cm 0.0cm]{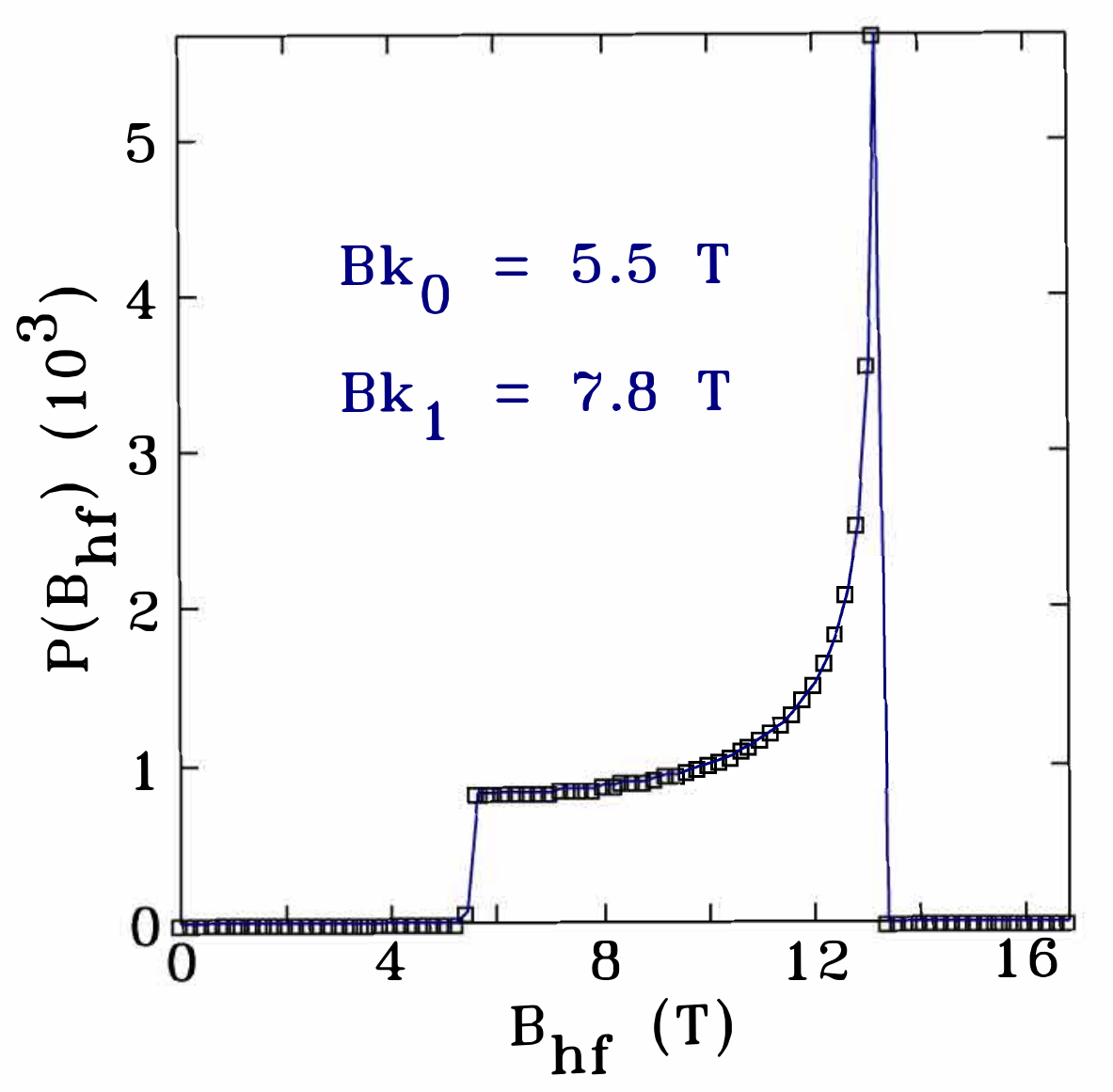}
		\caption {\textbf{Histogram for fitting the $\bm{T=16}$~K Mossb\"{a}uer spectrum.}  $B_{\text{hf}}$ is the hyperfine field and $P(B_{\text{hf}})$ is the probability distribution of $B_{\text{hf}}$.}
		\label{Mossbauer_Four_Prob}
			\end{figure}

The fitting sequence is as follows: Starting with an initial set of $Bk_n$, the evolution of $B_{\text{hf}}$ along $\mathbf{k}$ is evaluated as shown in  Supplementary Figure~\ref{Mossbauer_Bhf_kx}. From $B_{\text{hf}}$ versus $kx$, the histogram in  Supplementary Figure~\ref{Mossbauer_Four_Prob} is constructed, and the calculated $^{151}$Eu M\"ossbauer spectrum is obtained through a sum of magnetic patterns weighted according to the histogram. A conventional non-linear least-squares minimization routine is then used to adjust $Bk_n$, a uniform baseline, an overall scale factor, and the isomer shift. It is essential to note that since the modulation is explicitly assumed to be incommensurate with the crystal lattice, positions along $kx$ do not correspond to specific positions in the chemical unit cell. Rather, they represent how $B_{\text{hf}}$ is sampled by the Eu. As the periodicity of the modulation and the crystal cell are not related by a simple rational fraction, neighboring points along the $kx$ axis will be far apart in real space.

%\subsection{Results}
The room temperature $^{151}$Eu M\"ossbauer spectrum has a single peak much like the $T=20$~K spectrum shown in  Supplementary Figure~\ref{Mossbauer_spectra}. Fitting the spectrum gives an isomer shift of $-11.35(2)$~mm/s, indicating Eu$^{2+}$. No trivalent impurity (within a sensitivity of $1$\%) was detected. The similar isomer shifts at $20$~K ($-11.28(1)$~mm/s) and $5$~K ($-11.20(2)$~mm/s) confirm that the valence does not change over the temperature range studied. The fitted $B_{\text{hf}}$ of $26.46(5)$~T at $5$~K is typical of a fully ordered Eu$^{2+}$ moment \cite{Ref_13}, and the sharp lines are consistent with the Eu being in a single magnetic environment at $5$~K. The trigonal $\bar{3}m.$ point symmetry of the Eu site leads to the requirement of an axially-symmetric electric-field-gradient tensor, so the measurements are not sensitive to rotations of $\bm{\mu}$ about $\mathbf{c}$.

%		\begin{figure}[] %%% Mossbauer_Fourier T dependence%%%
%		\centering
%		\includegraphics[width=0.7\textwidth,trim=0.0cm 0.0cm 0.0cm 0.0cm]{Fourier_Bk0Bk1.pdf}
%		\caption {\textbf{Temperature dependence of the lowest-order Fourier components of $B_{\text{hf}}$ used to fit the spectra for $9~K<T\leq18$~K using the modulated model.} $Bk_0$ is a uniform dc offset and reflects non-modulated ordering of the Eu$^{2+}$ moments. $Bk_1$ is the fundamental harmonic component in the Fourier series describing the field distribution that arises from the incommensurate modulated order. The average field ($B_{\text{avg}}$) is also shown and provides a measure of the average ordered europium moment. The dashed line through $B_{\text{avg}}(T)$ is a fit to the $J$=$\frac{7}{2}$ Brillouin function yielding an ordering temperature of $17.5(1)$~K, while the same fit (dotted line) through $Bk_0(T)$ yields a transition temperature of $15.9(1)$~K. The dashed line through $Bk_1(T)$ is a guide to the eye.  The suppression of $Bk_1(T)$ upon cooling indicates that no modulation of $\mu$ exists at $T=6$~K.  }
%		\label{Mossbauer_Fourier}
%			\end{figure}
			
 Supplementary Figure~\ref{Mossbauer_spectra} shows that increasing the temperature not only leads to a gradual reduction in $B_{\text{hf}}$, but also to a clear increase in the linewidth. This broadening reflects a distribution of environments for the Eu, and may arise from either dynamic effects (e.g.\ slow paramagnetic relaxation) or from a static distribution of hyperfine fields. The former is inconsistent with both the observed evolution of the spectral shapes and with the continued observation of well-defined magnetic Bragg peaks in the neutron diffraction data. This leaves a static distribution of hyperfine fields as the source of the line broadening. As Eu$^{2+}$ is the only magnetic species present, a distribution in $B_{\text{hf}}$ necessarily reflects a distribution of moment magnitudes $\mu$.  We therefore turn to the distribution model given by Supplementary Equation~(\ref{eqn:fourier}) that has been successful in describing the order in a number Eu- and Fe- based compounds \cite{Ref_10,Ref_11,Ref_12,Ref_14}. The solid lines through the spectra shown in  Supplementary Figure~\ref{Mossbauer_spectra} for $9<T\leq18$~K demonstrate that this model provides an excellent description of the spectra.

Looking at  Supplementary Figure~\ref{characterization_1}d, for $T=6$~K only the uniform term is needed and $B_{\text{avg}}$ is equivalent to $Bk_0$. Thus, at $6$~K no modulation of $\mu$ is present, in agreement with the magnetic structure determined by neutron diffraction. However, by $11$~K a weak modulation is needed to fit the spectrum: $Bk_1$ is no longer zero and $Bk_0$ starts to fall below $B_{\text{avg}}$. $Bk_1(T)$ peaks at $\approx16$~K and $Bk_0$ is only lost by $18$~K. Fitting $B_{avg}(T)$ to a $J=\frac{7}{2}$ Brillouin function (appropriate for Eu$^{2+}$) yields $17.5(1)$~K, while doing the same for $Bk_0(T)$ gives $15.9(1)$~K, which is also where $Bk_1(T)$ peaks.  These two ordering temperatures are in good agreement with $T_{\text{N}1}=17.6(2)$~K and $T_{\text{N}2}=16.2(1)$~K. determined by neutron diffraction.  We also remark that $Bk_1(T)$ is finite proximate to the AF phase transitions, suggesting that the inferred modulation in $\mu$ may be related to the critical region around these phase transitions.

To summarize, our M\"ossbauer measurements indicate that the initial AF order that develops on cooling through $17.5$~K is dominated by an incommensurate, sine-like, modulation of $\mu$ along the direction of the AF propagation vector, but a significant uniform contribution is likely also present. On further cooling, both contributions grow, but at $16$~K the modulation starts to decline and the uniform term quickly dominates, until by $9$~K only order with a uniform (fixed-size) value of $\mu$ remains.

\section{S\lowercase{upplementary} N\lowercase{ote 4.}  Additional density functional theory results}
	\begin{figure}[] %%% DFT%%%
	\centering
	\includegraphics[width=0.7\textwidth,trim=0.0cm 0.0cm 0.0cm 0.0cm]{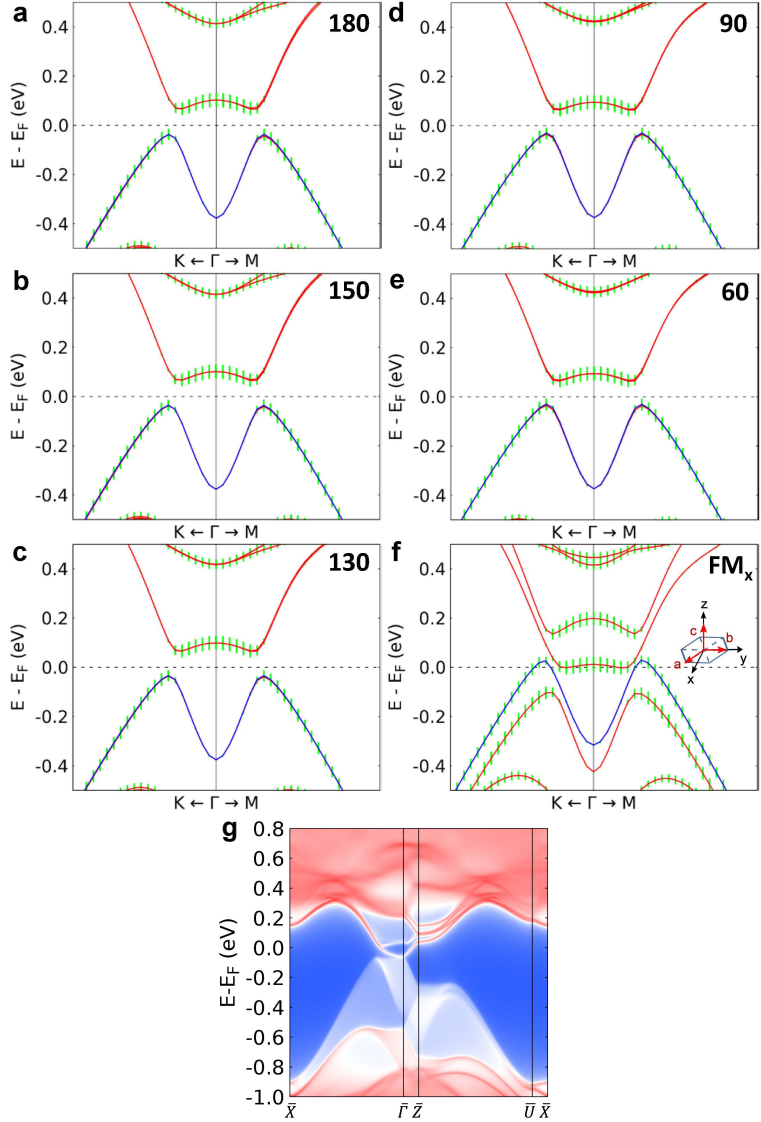}
	\caption {\textbf{Further results from density-functional-theory calculations including spin-orbit coupling.}  \textbf{a}--\textbf{e}, Results from calculations made after adiabatically changing the turn angle $\phi_{\text{rb}}$ associated with the broken-helix ground state. The value for $\phi_{\text{rb}}$ is indicated in each panel by the top-right number. The top valence band according to simple band filling is in blue and the rest of the bands are in red. Green vertical bars show As $4p_z$ orbital character and indicate band inversion near the minimal gap. \textbf{f}, Results for a ferromagnetic ground state with moments oriented along $\mathbf{x}$.  The relationship of $\mathbf{x}$ to the lattice is shown in the included diagram. \textbf{g}, Full view of the surface band structure and band gap for the $(110)$ projected surface along high-symmetry lines for $\phi_{\text{rb}} =130\degree$.} \label{DFT_SI}
\end{figure}

Density functional theory (DFT) calculations including spin-orbit coupling were made after adiabatically changing $\phi_{\text{rb}}$, in accordance with the magnetic space group constraints $\phi_{\text{rr}} + 2\phi_{\text{rb}} =
180\degree$ and blue moments alternately pointing along $\pm[\bar{1}10]$, to values spanning the region between pure $60\degree$-helix and A-type order shown in Fig.~\ref{fig_bloc}a of the main text.  Results of these calculations are shown in Figs.~\ref{band_structure}b--\ref{band_structure}d and Supplementary Figures~\ref{DFT_SI}a--\ref{DFT_SI}e, where the value of $\phi_{\text{rb}}$ for each panel is indicated by the number at the top right.  As described in the main text, band-inversion and the electronic-band gap remain over the whole $60\degree \leq \phi_{\text{rb}} \leq 180\degree$ region which implies that $\theta=\pi$ over the whole region.   Supplementary Figure~\ref{DFT_SI}f shows that for ferromagnetic order with the ordered moment lying within the basal plane band inversion remains but degeneracy is lifted and a Fermi surface forms.  A $12\times12\times3$ $k$-point mesh was used for the ferromagnetic calculation. Supplementary Figure~\ref{DFT_SI}g displays a full view of the surface band structure for the $(110)$ projected surface along high-symmetry lines for $\phi_{\text{rb}} =130\degree$. This extended view complements Figs.~\ref{band_structure}e and \ref{band_structure}f, revealing both the gap at $E_\text{F}$ and the surface Dirac cone at $E_\text{F}-0.061$~eV.

\section{S\lowercase{upplementary} N\lowercase{ote 5.}  Symmetry analysis of the magnetic phases}\label{Supp_Note_5}

In this section, we provide a detailed symmetry analysis of the magnetic space groups (MSGs) for the two observed magnetic phases: (i) the pure $60\degree$-helix order emerging upon cooling through $T_{\text{N}1} = 17.6(2)$~K, and (ii) the broken-helix order existing for $T\leq T_{\text{N}2}$. The material crystallizes according to the crystallographic space group (SG) $P6_3/mmc$ (No.\ $194$) and is built of hexagonal layers of Eu alternating with blocks of In$_2$As$_2$ along the crystallographic $\mathbf{c}$ axis~\cite{Ref_3,Ref_4}, see Fig.~\ref{Diff}c. The unit cell contains two Eu layers separated by a distance of $c/2$. In this section, the $\mathbf{c}$ axis will also be denoted as $\hat{c} \equiv [001]$ or $\hat{z}$, and in Supplementary Figure~\ref{fig:symmetry}a we depict our choice of $\hat{x}$, $\hat{y}$, and $\hat{z}$ axes with respect to the in-plane hexagonal directions $\hat{a} \equiv [100]$ and $\hat{b} \equiv [010]$.

The crystallographic space group $P6_3/mmc$ (No.\ $194)$ is generated by the following elements: inversion ($\mathcal{I}$), a $2\pi/3$ rotation around $\hat{z}$ ($C_{3, [001]}$), a $\pi$ rotation around $\hat{z}$ followed by a $\frac{c}{2}$  translation along $\hat{z}$  ($\{\left.C_{2, [001]}\right| 00\frac{c}{2}\}$), and a $\pi$ rotation around $[100]$: $C_{2,[100]}$. In the paramagnetic phase, the MSG is $P6_3/mmc1^{\prime}$ (No.\ $194.264$), which is a grey group containing time-reversal [$\mathcal{T}$ (also indicated by the prime symbol)] combined with any symmetry operation of the crystallographic SG.

The two helical magnetically ordered phases are described by a magnetic unit cell that is three times larger along $\hat{c}$ than the crystallographic unit cell, $c_{\text{mag}}=3c$. As a result, the two additional translations $ \hat{c}$ and $2\hat{c}$, which are $\hat{c}_{\text{mag}}/3$ and $2\hat{c}_{\text{mag}}/3$ with respect to the magnetic unit cell, need to be considered. Combined with the crystallographic SG, this leads to additional symmetry operations which are non-symmorphic with respect to the magnetic unit cell. Moreover, while  $\mathcal{T}$ by itself is no longer a symmetry element of the magnetically ordered phase, $\mathcal{T}$ combined with some spatial operations of the crystallographic SG are symmetry elements of the magnetic phase, and the MSG is of black-white type.  The MSG is defined by the subset of operations generated by the generators of the paramagnetic MSG $P6_3/mmc1^{\prime}$ combined with the  translations $\{\hat{c} = \hat{c}_\text{mag}/3, 2 \hat{c} = \hat{c}_{\text{mag}}/3 \}$ which leave the magnetic order invariant.

%\subsection{Pure $60\degree$-helix magnetic order}  
We first analyze the symmetries of the pure $60\degree$-helix phase that appears upon cooling through $T_{\text{N1}} = 17.6(2)$~K as shown in Fig.~\ref{fig_bloc}b in the main text. The ordered magnetic moments $\bm{\mu}$ are ferromagnetically aligned in each Eu layer. As shown in  Supplementary Figure~\ref{fig:symmetry}, neighboring layers are oriented at an angle of $\pm 60\degree$ to each other such that the structure describes a pure $60\degree$ helix propagating along $\hat{z}$. This magnetic structure repeats every six Eu-layers, leading to a tripling of the chemical unit cell. The ordered moments in layers $3$ and $6$ point along the high-symmetry directions $[\bar{1} 1 0]$ and $[1 \bar{1} 0]$ (denoted in blue), respectively. We verified that there are $12$ magnetic symmetry operations that leave the structure invariant. For a helix that rotates counterclockwise around $\hat{c}$, the corresponding MSG is $P6_12^{\prime}2^{\prime}$ (No.\ $178.159$), which is generated by the set

\begin{equation}
\left\{\mathbbm{1},\bigl\{\left.C_{2,[001]}\right| 00\frac{c_{\text{mag}}}{2}\bigr\},C_{3,[001]}\bigl\{\left.\mathbbm{1}\right|00\frac{c_{mag}}{3}\bigr\},\mathcal{T}C_{2,[100]}\right\}\text{ ,}
\label{MSG_perfect_helix}
\end{equation}

\noindent where $\mathbbm{1}$ is the identity element. Note that a helix with opposite helicity (i.e.\ clockwise rotation around $\hat{c}$) is described by MSG $P6_52^{\prime}2^{\prime}$ (No.\ $179.165$).  

The symmetry of the pure $60\degree$-helix order can be readily checked graphically, as shown in Supplementary Figure~\ref{fig:symmetry}. It is important to note that a rotation around an  axis different from $[001]$ leads to a \textit{reshuffling} of the Eu-layers as shown in Supplementary Figures~\ref{fig:symmetry}b and \ref{fig:symmetry}c for $C_{2, [100]}$. The reshuffling depends on the choice of origin of the coordinate system. Here, we choose the origin to lie in Eu layer $2$. Whereas the form of the generators depends on the choice of origin, the resulting MSG [$P6_12^{\prime}2^{\prime}$ (No.\ $178.159$)] is independent of the origin choice.

Importantly, these MSGs contain the element $\mathcal{T}C_{2, [100]}$, which reverses an odd number of space-time dimensions, and therefore leads to a quantization of the magnetoelectric coupling (axion) angle $\theta$.  Further, the magnetic symmetry $\mathcal{T}C_{2, [100]}$ protects exotic unpinned gapless surface Dirac states on the $(2\bar{1}0)$ surface ~\cite{Ref_15}. This MSG contains five more $2'= \mathcal{T} C_{2}=$ elements that protect unpinned Dirac cones on five more surfaces. These are the surfaces $(\bar{1} 2 0)$ and $(110)$, which are related to $(2\bar{1}0)$ by $C_{3,[001]}$ and $C^2_{3,[001]}$ rotations, respectively, as well as the surfaces $(100)$, $(010)$ and $(\bar{1} 1 0)$, which are related to each other by $C_{3,[001]}$ rotations. A list of the $2^{\prime}$ symmetry axes and surfaces with gapless Dirac cones for the various magnetic phases is given in Fig.~\ref{Field_Pol}c.

%\subsection{Broken-helix magnetic order} 
In the broken-helix state occurring for $T \leq T_{\text{N2}} = 16.2(1)$~K, the (blue) ordered magnetic moments in Eu-layers $3$ and $6$ continue to point along the high-symmetry directions $[\bar{1}10]$ and $[1\bar{1}0]$, respectively. The red ordered moments (red layers), however, are not oriented at $\pm60\degree$ with respect to the layer immediately above or below as shown in Figs.~\ref{Diff}d, \ref{fig_bloc}a, and Supplementary Figure~\ref{bH_order}. As a result, $C_{3, [001]}$, which is a symmetry element for the $60\degree$-helix order, is lost in the broken-helix state. The broken-helix phase is described by MSG $C2^{\prime}2^{\prime}2_1$ (No.\ $20.33$), which is generated by the element set
\begin{equation}
\{\mathbbm{1},\bigl\{\left.C_{2, [001]}\right| 00\frac{c_{\text{mag}}}{2}\bigr\},\mathcal{T}C_{2,[110]}\} \text{ .}
\label{MSG_broken_helix}
\end{equation} 
Here, we have set the origin to lie in  Eu layer $3$ (blue). Since the MSG contains the element $\mathcal{T} C_{2[110]}$, $\theta$ is still quantized. The $[110]$ direction is perpendicular to the high-symmetry direction of the blue moments. Note that, in contrast to the $60\degree$-helix order, here we only have two $2^{\prime}$ operations, one $2^{\prime}$ axis along $[110]$ and the other along $[1\bar{1}0]$. This has profound consequences on the location of the unpinned gapless surface Dirac cones protected by $2^{\prime}$ \cite{Ref_15}: they appear only on $(110)$ and $(\bar{1}10)$ surfaces for broken-helix magnetic order, whereas they can also be found on four other surfaces [$(100)$, $(010)$, $(\bar{2}10)$ and $(\bar{1}20)$] for the pure $60\degree$ helix state. 

\begin{figure}[]
\centering
\includegraphics[width=0.8\textwidth]{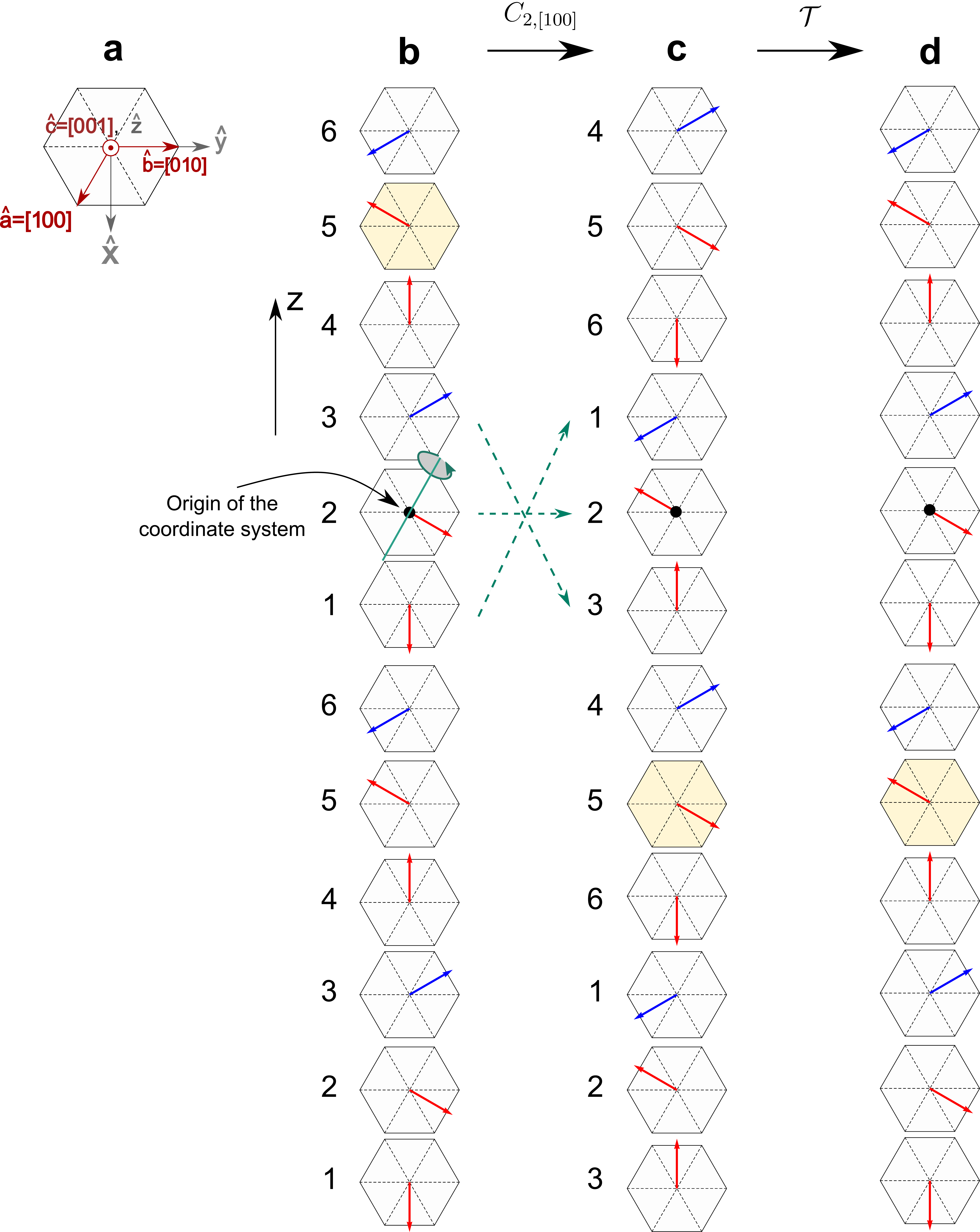}
\caption{\textbf{Action of $\bm{\mathcal{T}C_{2,[100]}}$ on pure $\bm{60\degree}$-helix order.} \textbf{a}, Crystallographic $\hat{a}$, $\hat{b}$ and $\hat{c}$ axes for hexagonal EuIn$_2$As$_2$ together with our choice of $\hat{x}$, $\hat{y}$ and $\hat{z}$ axes. \textbf{b}, $60\degree$-helix order where two magnetic unit cells are shown. \textbf{c, d}, The remaining panels show the action of $\mathcal{T}C_{2,[100]}$ on the $60\degree$-helix phase:  a $\pi$ rotation around the $[100]$ axis (c) followed by a time-reversal operation (d) leaves the initial structure invariant. Therefore, $\mathcal{T}C_{2,[100]}$ is a symmetry operation of the pure $60\degree$-helix order. One of the Eu-layers is highlighted to facilitate the visualization of the action of each of the space-time transformations. Note that we set the origin of the coordinate system in the second Eu layer.}
\label{fig:symmetry}
\end{figure}

\begin{figure}[]
\centering
\includegraphics[width=0.69\textwidth]{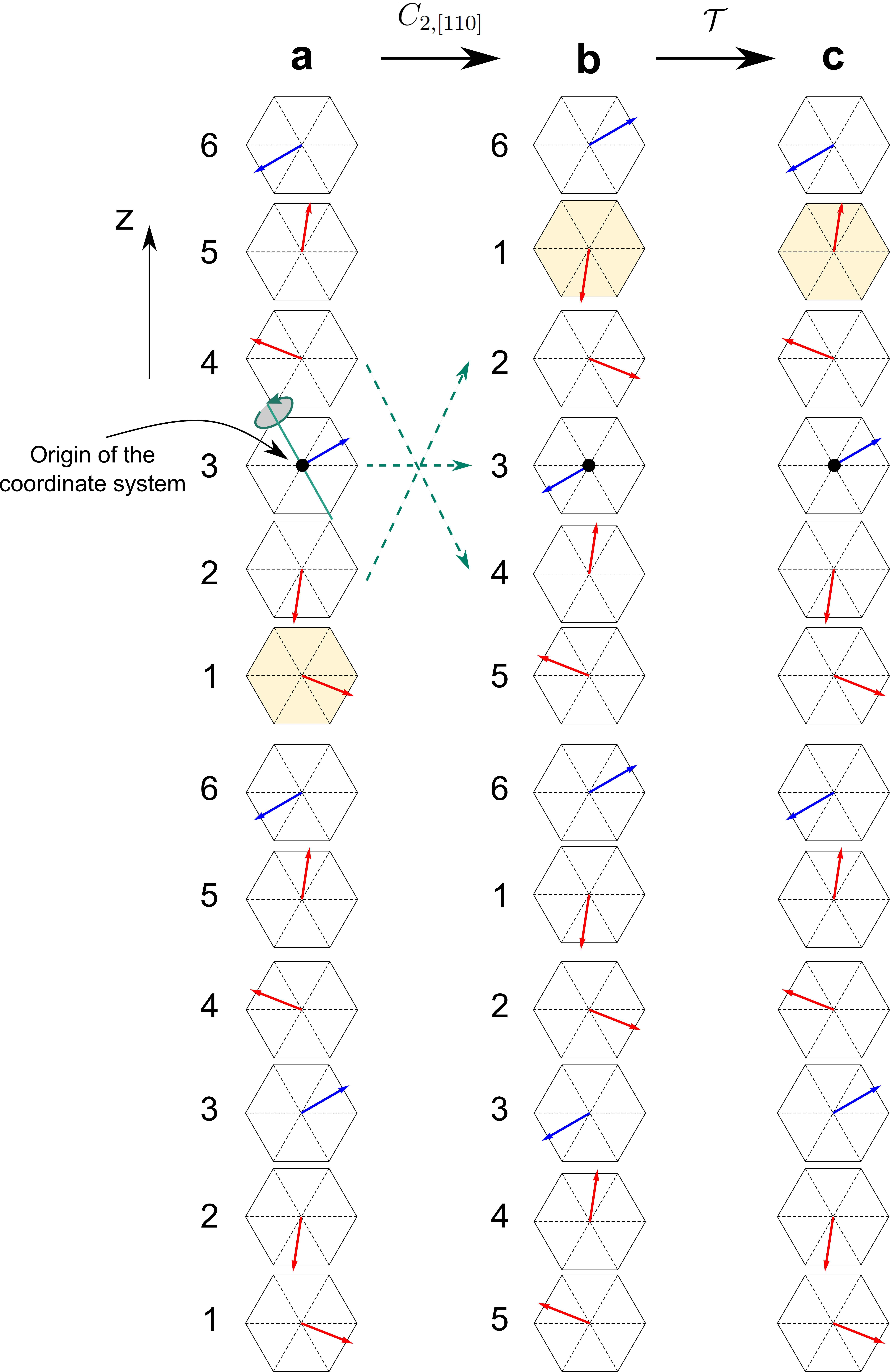}
\caption{\textbf{Action of $\bm{\mathcal{T}C_{2,[110]}}$ on broken-helix order.} \textbf{a}, Initial magnetic structure, where two magnetic unit cells are shown. \textbf{b, c}, A  $\pi$ rotation around the $[110]$ axis (b) followed by a time-reversal operation (c) leaves the initial structure invariant. Therefore, $\mathcal{T}C_{2,[110]}$ is a symmetry operation of the broken-helix structure. Similar to Supplementary Figure~\ref{fig:symmetry}, one of the Eu-layers is highlighted to facilitate the visualization of the action of each of the space-time transformations. Here, the origin of the coordinate system is set in the third Eu layer.} \label{fig:symmetry_broken}
\end{figure}

% subsection broken_helix_magnetic_order (end)

%\subsection{Collinear A-type magnetic order} 
We next consider A-type antiferromagnetic order similar to that discussed in Supplementary Reference~[\onlinecite{Ref_16}] for MSG $Cmcm$ (No.\ $63.457$) and ordered moments lying along $[100]$. Here, the ordered Eu magnetic moments are again ferromagnetically-aligned in each Eu-layer, but the stacking along $\hat{c}$ is simply antiferromagnetic. In contrast to Supplementary Reference~[\onlinecite{Ref_16}], we consider moments lying parallel to the high-symmetry $[\bar{1}10]$ directions and use MSG $Cm^{\prime}c^{\prime}m$. The A-type order is smoothly connected to the experimentally determined broken-helix and pure $60\degree$-helix orders by varying the helical-turn angles $\phi_{\text{rb}}$ and $\phi_{\text{rr}}$, where $\phi_{\text{rr}}= \pi - 2 \phi_{\text{rb}}$\, as shown in Fig.~\ref{fig_bloc}a of the main text. $|\phi_{\text{rb}}| = |\phi_{\text{rr}}| = \pi$ for the A-type state. Since the crystallographic unit cell contains two Eu layers, the magnetic unit cell is equal to the crystallographic one in this case. The MSG of A-type order with ordered moments along $[1\bar{1}0]$ is $Cm^{\prime}c^{\prime}m$ (No.\ $63.462$), which is generated by
\begin{equation}
\{\mathbbm{1},\mathcal{I},\bigl\{\left.C_{2,[001]}\right|00\frac{c}{2}\bigr\},\mathcal{T}C_{2,[110]}\}\,.
\label{eq:AFM}
\end{equation}
We find that gapless surface Dirac cones, which can be shifted away from the center of the surface Brillouin zone, occur on the $(110)$ and $(\bar{1}10)$ surfaces. In contrast to the $60\degree$-helix order and the broken-helix order, here the Dirac cones are constrained to occur along the $\bar{\Gamma}$-$\bar{\mathrm{X}}$ direction of the surface Brillouin zones due to the mirror symmetry $\bigl\{\left.m_{[001]}\right|00\frac{c}{2}\bigr\}$ that is preserved on both surfaces.

Importantly, due to the presence of $\mathcal{I}$ the energy bands have well-defined parity. This allows us to easily infer the value of  $\theta$ by analyzing the parity-based $\mathbb{Z}_4$ invariant, as explained in the main text. By our DFT calculations, we find $\mathbb{Z}_4 = 2$ for the A-type state, which corresponds to $\theta = \pi$ and an axion insulator (AXI) phase. As our DFT results show that the band gap never closes when we continuously change the angles $\phi_{rb}$ and $\phi_{rr}$ from $|\phi_{\text{rr}}| = |\phi_{\text{rb}}| = \pi$ to their experimentally determined values in the broken- and $60\degree$-helix states, we conclude that both the broken- and pure $60\degree$-helix states also have $\theta = \pi$ and are therefore AXIs. Note that while $\mathcal{I}$  is absent in the experimentally determined helical structures, the combination $\mathcal{T}C_{2}=2^{\prime}$ guarantees quantization of $\theta$, as it reverses an odd number of space-time coordinates. Finally, we note that, as discussed in Supplementary Reference~[\onlinecite{Ref_16}], A-type order with ordered moments along $[100]$ or $[010]$ is described by MSG $Cmcm$ (No.\ $63.457$), whereas A-type order with moments along $[001]$ is described by MSG $P6_{3^{\prime}}/m^{\prime} m^{\prime}c$ (No.\ $194.268$).

%\subsection{Magnetic phase transitions and magnetic subgroup relations} 
Several magnetic phase transitions occur in EuIn$_2$As$_2$ upon cooling, which in Landau theory for second-order phase transitions may be related to transitions from a symmetry group to one of its subgroups. Starting from the high-temperature paramagnetic phase, the compound undergoes a transition to pure $60\degree$-helix order below $T_{\text{N}1}=17.6(2)$~K, followed by a second transition to broken-helix order below $T_{\text{N}2}=16.2(1)$~K. From the point of view of group-subgroup relations, there are three different paths the system could follow between the high-temperature MSG $P6_3/mmc1^{\prime}$ and the low-temperature MSG $C2^{\prime}2^{\prime}2_{1}$ as shown in  Supplementary Figure~\ref{MSG_tree}. The left path in  Supplementary Figure~\ref{MSG_tree} corresponds to the Eu moments first aligning ferromagnetically along $[001]$, followed by a transition to A-type order, and finally by a third transition to the broken-helix phase. This path is ruled out by our neutron diffraction data, since no out-of-plane component of the ordered magnetic moment is observed (see  Supplementary Figure~\ref{100_order_para}) and only two ordering transitions were seen.

The other two paths in Supplementary Figure~\ref{MSG_tree} correspond to transitions from the paramagnetic state to pure $60\degree$-helix order and, finally, to the broken-helix state. Based on our experiments, both of these routes are possible with the only difference between the two cases being the handedness (chirality) of the pure $60\degree$ helix phases: the moments rotate clockwise around $[001]$ for the center path and counterclockwise around $[001]$ for the right path. It is important to note that the handedness of the helical states cannot be resolved from our data, and it is likely that domains with different handedness coexist in the sample. Also, note that the broken-helix state also exhibits a handedness, yet both right- and left-handed chiralities are described by MSG $C2^{\prime}2^{\prime}2_{1}$.

\begin{figure}[] %%% MSG Tree%%%
\centering
\includegraphics[width=0.5\textwidth]{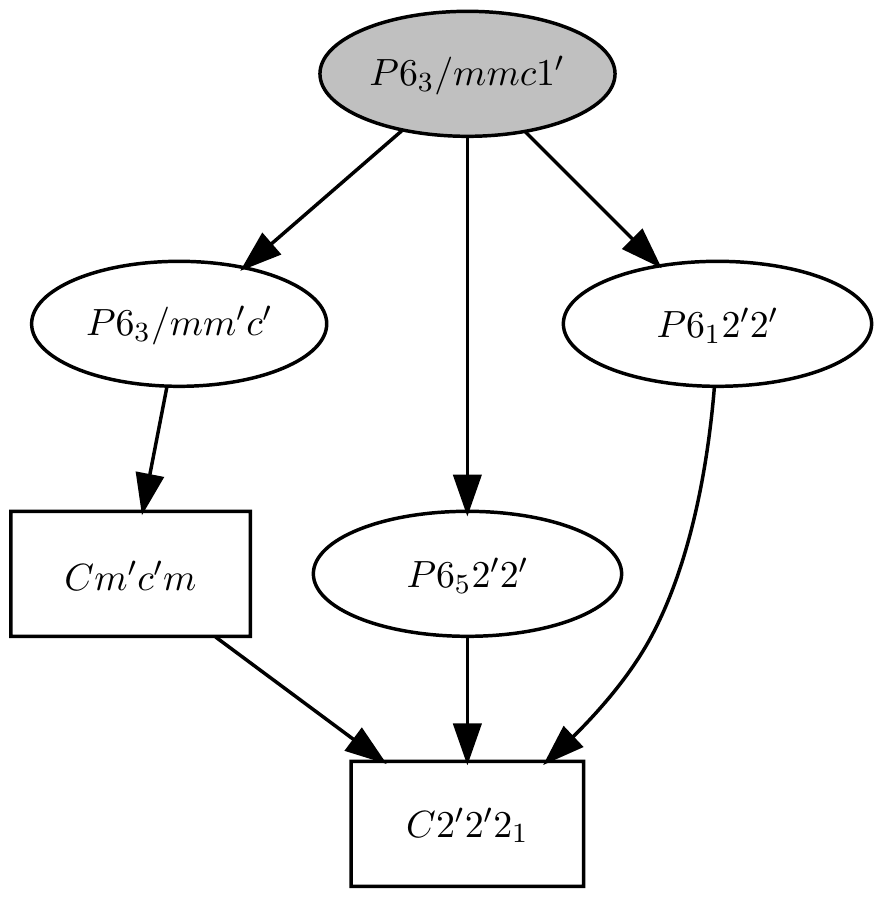}
\caption{\textbf{Magnetic subgroup tree.} Magnetic group--subgroup tree connecting the paramagnetic magnetic space group (MSG) $P6_3/mmc1^{\prime}$ to MSG $C2^{\prime}2^{\prime}2_{1}$, which corresponds to the  broken-helix order. The middle and right paths reach the low-temperature phase via intermediate $60\degree$ helical order propagating along $\mathbf{c}$ with either counterclockwise (right path) or clockwise (center path) chirality. Both of these paths agree with the experimental results. The left path traverses $A$-type order ($Cm^{\prime}c^{\prime}m$) with moments along $[\bar{1}10]$ (or $[1\bar{1}0]$), which is not experimentally realized. This diagram is taken from Supplementary Reference~\cite{Ref_1}.}
\label{MSG_tree}
\end{figure}

%\subsection{Field-polarized magnetic order} 
 Polarizing the Eu moments along a specific direction via applying an external magnetic field offers tuning of the topological properties of the material. As shown in Fig.~\ref{Field_Pol}b and Supplementary Figure~\ref{characterization_2}b, at $T=5$~K a magnetic field of $\mu_{0}H\approx1$ to $2$~T aligns the Eu magnetic moments parallel to the field, which corresponds to a field-polarized state. Below, we study the symmetry and topological consequences of a field-polarized state by considering three orientations for the external field.

%\subsubsection{Perfectly-polarized phase with moments along $[\bar{1}10]$}\label{sub:pol-110}
We first consider a magnetic field along the high-symmetry direction $[\bar{1}10]$, which is the direction that the blue Eu ordered magnetic moments lie along. The resulting field-polarized state with all moments aligned along $[\bar{1}10]$ is characterized by MSG $Cm^{\prime}cm^{\prime}$ (No.\ $63.464$), which is generated by
\begin{equation}
    \{\mathbbm{1}, \mathcal{I},\mathcal{T}\bigl\{\left.C_{2,[001]}\right|00\frac{c}{2}\bigr\},\mathcal{T}C_{2,[110]}\} \text{ .}
    \label{eq:FM_blue}
\end{equation}

\noindent In this phase, there is both $\mathcal{T}C_{2,[110]}\equiv 2^{\prime}_{[110]}$ and $\mathcal{T}\bigl\{\left.C_{2,[001]}\right|00\frac{c}{2}\bigr\} \equiv\bigl\{\left.2^{\prime}_{[001]}\right|00\frac{c}{2}\bigr\}$. While the former protects a gapless Dirac cone on the $(110)$ surface, the latter is a screw axis that is naturally broken on the $(001)$ surface. Therefore, $(110)$ is the only surface hosting an exotic surface state in this polarized phase. Importantly, because the mirror symmetry $\bigl\{\left.m_{[\bar{1}10]}\right|00\frac{c}{2}\bigr\}$ is preserved in this surface, the Dirac cone occurs along the $\bar{\Gamma}$-$\bar{\mathrm{Z}}$ direction in the surface Brillouin zone.

%\subsubsection{Perfectly-polarized phase with moments along $[100]$}
Another possibility is to align the Eu moments along  $[100]$. In this case, the MSG is $Cmc^{\prime}m^{\prime}$ (No.\ $63.463$), with generators
\begin{equation}
    \left\{\mathbbm{1}, \mathcal{I}, C_{2,[100]}, \mathcal{T}\bigl\{\left.C_{2,[001]}\right|00\frac{c}{2}\bigr\}\right\} \text{ .}
    \label{eq:generators100}
\end{equation}

\noindent The combinations of the generators in Supplementary Equation~(\ref{eq:generators100}) result in two $2^{\prime}$ operations:  $\bigl\{\left.2^{\prime}_{[001]}\right|00\frac{c}{2}\bigr\} $ and  $ \bigl\{\left.2^{\prime}_{[120]}\right|00\frac{c}{2}\bigr\}$. As above, the screw operation $\bigl\{\left.2^{\prime}_{[001]}\right|00\frac{c}{2}\bigr\} $ is naturally broken on the $(001)$ surface. Therefore, here we only have a Dirac surface state on the $(010)$ surface, which is pinned along the $\bar{\Gamma}$-$\bar{\mathrm{Z}}$ direction due to an additional $m_{[100]}$ mirror symmetry. 

%\subsubsection{Perfectly-polarized phase with moments along $[001]$}
If we apply a magnetic field perpendicular to the Eu layers so that the moments align parallel to $\hat{c}$, we obtain a magnetic state belonging to MSG $P6_3/mm^{\prime}c^{\prime}$ (No.\ $194.270$). The generators of this group are
\begin{equation}
    \{\mathbbm{1},\mathcal{I},C_{3,[001]},\bigl\{\left.C_{2,[001]}\right|00\frac{c}{2}\bigr\},\mathcal{T}C_{2,[100]}\} \text{\,.}
\end{equation}

\noindent They give rise to $24$ symmetry operations, among which six are combinations of $\mathcal{T}$ and $C_{2}$ rotations. These $2^{\prime}$ operations can be grouped into two different classes: one class contains the element $2^{\prime}_{[100]}$ and those related to it by $C_{3,[001]}$. As a consequence, Dirac surface states emerge on the surfaces $(2\bar{1}0)$, $(1\bar{2}0)$ and $(110)$.  The second class contains the element $\bigl\{\left.2^{\prime}_{[120]}\right|00\frac{c}{2}\bigr\}$ and those related to it by $C_{3,[001]}$. Note that the direction of these $2^{\prime}$ operations are normal to the side faces of the hexagonal crystallographic unit cell, and they protect Dirac surface states on the $(100)$, $(010)$ and $(1\bar{1}0)$ surfaces.

Note that the surfaces hosting exotic states are the same as those for pure $60\degree$-helix order. The difference here is that the Dirac cones are no longer completely unpinned, but are constrained to occur along $\bar{\Gamma}$-$\bar{\mathrm{X}}$ because of the additional mirror symmetry $\bigl\{\left. m_{[001]}\right|00\frac{c}{2}\bigr\}$ resulting from the combination of inversion $\mathcal{I}$ and the screw operation $\bigl\{\left.C_{2,[001]}\right|00\frac{c}{2}\bigr\}$.

\begin{table}[]
\centering

\begin{tabular}{|l|c|c|}
\hline
                                            & MSG                        & \multicolumn{1}{l|}{\; $X_{BS}$\; } \\ \hline
Perfect helix  \;                             & \;$P6_12^{\prime}2^{\prime}$ (No.\ $178.159$) \;  & $1$                             \\ \hline
Broken helix \;                                & \;$C2^{\prime}2^{\prime}2_{1}$ (No.\ $20.33$)\;  & $1$                             \\ \hline
A-type $\parallel \left[\bar{1}10\right]$ \; & \;$Cm^{\prime}c^{\prime}m$ (No.\ $63.462$)\;        & $2$                             \\ \hline
Field-pol. $\parallel \left[\bar{1}10\right]$ \;     & \;$Cm^{\prime}cm^{\prime}$ (No.\ $63.464$)\;        & $(2,2)$                         \\ \hline
Field-pol. $\parallel [100]$ (or $[010]$) \;       & \;$Cmc^{\prime}m^{\prime}$ (No.\ $63.464$)\;        & $(2,2)$                         \\ \hline
Field-pol. $\parallel [001]$ \;                     & \;$P6_3/mm^{\prime}c^{\prime}$ (No.\ $194.270$) \; & $(3,6)$                         \\ \hline
\end{tabular}
\caption{\textbf{Symmetry-based indicators of band topology.} Data taken from Supplementary Reference~\onlinecite{Ref_17} for all of the magnetic phases described in this section.}
\label{tab:my-table}
\end{table}

%\subsection{Discussion}
It becomes evident from the previous analysis that we can use an external magnetic field to either induce or destroy topological states on different surfaces of EuIn$_2$As$_2$. For instance, applying a magnetic field along $[\bar{1}10]$ inside the pure $60\degree$-helix phase gaps out Dirac cones on $(100)$, $(010)$, $(2\bar{1}0)$ and $(1\bar{2}0)$ surfaces and pins the gapless Dirac cones on $(110)$ and $(1\bar{1}0)$ to the $\bar{\Gamma}$-$\bar{\mathrm{Z}}$ direction in the surface Brillouin zone. If, on the other hand, a field parallel to $[001]$ is applied in the pure $60\degree$-helix phase, the Dirac cones remain gapless on all six surfaces but are now all constrained to lie along the $\bar{\Gamma}$-$\bar{\mathrm{X}}$ direction. %The situation is reversed for the broken-helix and A-type orders. In these cases, it is a field along $[001]$ that destroys the Dirac surface states at $(110)$ surfaces and instead moves them to the $(100)$ surface.

In Supplementary Table~\ref{tab:my-table}, we show the symmetry-based indicators $X_{BS}$ of topology for the magnetic helical phases and the field-polarized phases. While symmetry indicators are absent for the MSGs describing the two helical phases, since $X_{BS} = 1$ means that there exist no symmetry indicators, the MSGs of A-type order and the field-polarized phases feature differing symmetry-based indicators.


\begin{thebibliography}{99}
\bibitem{Ref_20} $\bm{\tau}_2=(0,0,1)$ is equivalent to $\bm{\tau}_2=(0,0,0)$ for $P6_3/mmc$. We use $\bm{\tau}_2=(0,0,1)$ to facilitate presentation of the diffraction data.
\end{thebibliography}
\end{document}